\documentclass[11pt]{article}
\usepackage[utf8]{inputenc}
\usepackage{amsfonts}
\usepackage{amsmath}
\usepackage{amssymb}
\usepackage{indentfirst}
\usepackage{graphicx}
\usepackage[colorlinks]{hyperref}
\usepackage{cite}

\usepackage{graphicx}
\usepackage{latexsym}
\usepackage{bbold}
\usepackage{wasysym}
\usepackage{calligra}
\usepackage{float}
\usepackage{ulem}
\usepackage{inputenc}
\usepackage{xspace}
\usepackage{url}
\usepackage{epstopdf}
\usepackage{tikz}
\usepackage{amsthm}

\usepackage{inputenc}

\numberwithin{equation}{section}
\allowdisplaybreaks
\makeatletter
\newcommand{\beq} {\begin{equation}}
\newcommand{\eeq} {\end{equation}}
\newtheorem{prop}{Proposition}

\definecolor{blue-violet}{rgb}{0.54, 0.17, 0.89}
\definecolor{PineGreen}{cmyk}{0.92, 0, 0.59, 0.25}

\begin{document}

\begin{titlepage}
\vskip 4cm

\begin{center}
\textbf{\LARGE{{Parity Violating Metric-Affine Gravity Theories}}}
\par\end{center}{\LARGE \par}

\begin{center}
	\vspace{1cm}
	\textbf{Damianos Iosifidis}$^{\ast}$,
	\textbf{Lucrezia Ravera}$^{\dag,\star}$,
	\small
	\\[5mm]
	$^{\ast}$\textit{Institute of Theoretical Physics, Department of Physics Aristotle University of Thessaloniki, }\\
	\textit{ 54124 Thessaloniki, Greece.}
	\\[3mm]
    $^{\dag}$\textit{DISAT, Politecnico di Torino, Corso Duca degli Abruzzi 24, 10129 Torino, Italy.}
	\\[3mm]
	$^{\star}$\textit{INFN, Sezione di Torino, Via P. Giuria 1, 10125 Torino, Italy.} \\[5mm]
	\footnotesize
	\texttt{diosifid@auth.gr},
	\texttt{lucrezia.ravera@polito.it},
	\par\end{center}
\vskip 20pt
\centerline{{\bf Abstract}}
\medskip
\noindent

We study a parity violating Metric-{A}ffine gravitational theory given by the Einstein-Hilbert action plus the so-called Holst term in vacuum. We find out that for a certain value of the Barbero-Immirzi parameter the total action possesses a remarkable invariance under particular transformations of the affine connection. We prove that in all cases, with appropriate gauge choices, the connection reduces to the Levi-Civita one and that the theory turns out to be equivalent to general relativity in vacuum. Subsequently, we generalize our discussion and analyze the case of Metric-Affine $f(R)$ gravity plus the Holst term. In particular, we show that for $f'(R) \neq \text{constant}$ the theory results to be on-shell equivalent to a metric-compatible torsionless Scalar-Tensor model. Matter coupling of the aforementioned models is also discussed, together with explicit examples and applications.

\end{titlepage}

\newpage {\baselineskip=12pt \hypersetup{linkcolor=black}\tableofcontents{}}   

\noindent\rule{162mm}{0.4pt}

\section{Introduction}\label{introduction}

In the 19th century, the branches of Mathematics and Physics experienced an extraordinary
progress with the emergence of non-Euclidean geometry. In particular, the development of Riemannian
geometry led to {several} important results, among which the rigorous mathematical formulation {of general relativity} (GR). In spite of the great success and predictive power of {GR}, there are still some open issues whose understanding and solution may need the formulation {of new theoretical frameworks} as well as generalizations and extensions of Riemannian geometry. {{One} way to step beyond Riemannian geometry consists in {releasing} the Riemannian {assumptions} of metric comparability and torsionlessness of the connection and therefore allow for non-vanishing torsion and {nonmetricity} along with curvature. The latter constitutes a non-Riemannian geometry \cite{eisenhart2012non} which is the geometric arena where Metric-Affine Gravity {(MAG)} \cite{Hehl:1994ue} (see also \cite{Iosifidis:2019jgi} and references therein) theories are developed.}

In this generalized set-up, where geometrical objects have a clear physical meaning, the metric and the affine connection are not related a priori, and
a relation between them may be found only after having solved the equations of motion of the theory. Moreover, many modified theories of gravity can be obtained as special cases {of MAG}. MAG models, among which the peculiar case of $f(R)$ theories of gravity \cite{Sotiriou:2008rp},\footnote{{Here and in the sequel,} $f(R)$ denotes an arbitrary function of the scalar curvature $R=g^{\mu \nu} R_{\mu \nu}\left(\Gamma\right)$, with $\Gamma$ a general affine connection; see also \cite{Iosifidis:2019jgi} and references therein for details.} led to many relevant results and have been proved to have various applications in both Mathematics and Physics \cite{teyssandier1983cauchy,sotiriou2006f,Klemm:2018bil,Klemm:2020mfp,Klemm:2019izb,Klemm:2020gfm}. Even though the extensive study of MAG has started a few decades ago, there are many open issues that still need to be addressed, whose understanding could provide remarkable insights in the gravity context.

Driven by {the above motives}, in this paper we analyze in detail MAG theories in four (3+1) spacetime dimensions, providing new interesting results. In particular, we start by studying a parity violating MAG theory involving the Einstein-Hilbert (EH) term and the so-called Holst term \cite{Holst:1995pc} in vacuum. {The latter is defined by the contraction of the Riemann tensor with the Levi-Civita one and its presence on the gravitational sector is perfectly allowable\footnote{Indeed, just by dimensional analysis there is nothing preventing us to add this term. As a matter of fact, along with {the} Ricci scalar{,} these are the only two scalar {and pseudo-scalar} combinations one can write down that are linear in the Riemann tensor. Note that for Riemannian geometries the Holst term vanishes identically due of the absence of torsion in {that} case.} (if not mandatory){.}}
Here, let us mention that the addition of the Holst term in gravity theories was already considered in past literature in the presence of torsion (Einstein-Cartan gravity) \cite{Kazmierczak:2008iw,Banerjee:2010yn}, also at the quantum level \cite{Shapiro:2014kma}. 
In our analysis, involving both torsion and nonmetricity, we find that for a particular value of the Barbero-Immirzi (BI) parameter \cite{Barbero:1994ap,Immirzi:1996di,Geiller:2012dd} the total action possesses a remarkable invariance under peculiar transformations of the affine connection, which, to our knowledge, is reported {here} for the first time. {We prove that, in all cases,} with appropriate gauge choices, the connection reduces to the Levi-Civita one and that the theory turns out to be equivalent to vacuum GR. We also discuss matter coupling of the aforementioned model. Subsequently, we focus on another MAG action given by {Metric-Affine} $f(R)$ gravity plus the Holst contribution. Remarkably, we show that in the case $f'(R) \neq \text{constant}$ the theory results to be on-shell equivalent to a metric-compatible torsionless Scalar-Tensor model. We also discuss explicit examples with $f(R)=R+\varepsilon R^2$ {(where $\varepsilon$ is a constant parameter with dimensions of inverse mass squared)} and $f(R)=R^2$. For the latter case, we also provide an application to homogeneous cosmology and obtain {a solution}.

The paper is organized as follows: In Section \ref{prel} we give some preliminary theoretical background together with notation and conventions, which will be useful in the rest of the work. In Section \ref{EHHolst} we study the {Metric-Affine} theory given by the EH action plus the Holst term in vacuum.
In particular, {there} will be distinct cases to study separately: $\alpha \neq \pm \frac{i}{2}$ and  $\alpha=\pm \frac{i}{2}$, where $\alpha$ is the BI parameter. We will show that in the former case, {projective invariance of the action allows to} perform a proper gauge choice in such a way to end up with a final connection that is just Levi-Civita. In this case, as we will see, the theory results to be equivalent to GR in vacuum. On the other hand, we will show {that, remarkably,} in the case in which $\alpha=\pm \frac{i}{2}$ the total action enjoys an enlarged symmetry under affine connection transformations. {Also in this case, the theory reduces to vacuum GR and the final connection results to be the Levi-Civita one.}
Subsequently, in Section \ref{fRHolst} we study the MAG theory given by {Metric-Affine} $f(R)$ gravity plus the Holst term. Here, in particular, we find that in the case $f'(R) \neq \text{constant}$ the theory results to be on-shell equivalent to a metric-compatible torsionless Scalar-Tensor model. { Therefore, we {obtain and explicitly demonstrate} a remarkable result: Metric-Affine $f(R)$ gravity plus {the} Holst {term} is equivalent to a Scalar-Tensor {theory} with vanishing torsion and nonmetricity.} {Moreover, we} analyze explicit examples {together with} an application to homogeneous cosmology.
Matter coupling is also discussed in all cases. Finally, we conclude our work with a discussion and possible future developments.

\section{Preliminaries}\label{prel}

Let us start by giving, in the sequel, some theoretical background of {MAG theories}, reviewing the main geometric objects entering such models.
We adopt the same notation and conventions of \cite{Iosifidis:2019jgi} and consider {models in four (3+1) spacetime dimensions.} Our metric convention is $\eta= \text{diag} (-,+,+,+)$.
The generic decomposition of an affine connection reads
\begin{equation}\label{gendecompaffconn}
{\Gamma^\lambda}_{\mu\nu} = \tilde{\Gamma}^\lambda_{\phantom{\lambda}\mu\nu} + {N^\lambda}_{\mu\nu}\,,
\end{equation}
where the distortion tensor ${N^\lambda}_{\mu\nu}$ and the Levi-Civita connection 
$\tilde{\Gamma}^\lambda_{\phantom{\lambda}\mu\nu}$ are respectively given by
\begin{equation}\label{distortion}
{N^\lambda}_{\mu\nu} = \underbrace{\frac12 g^{\rho\lambda}\left(Q_{\mu\nu\rho} + Q_{\nu\rho\mu}
- Q_{\rho\mu\nu}\right)}_{\text{deflection}} - \underbrace{g^{\rho\lambda}\left(S_{\rho\mu\nu} +
S_{\rho\nu\mu} - S_{\mu\nu\rho}\right)}_{\text{contorsion} \, \equiv \, {K^\lambda}_{\mu \nu}}\,,
\end{equation}
\begin{equation}\label{lcconn}
\tilde{\Gamma}^\lambda_{\phantom{\lambda}\mu\nu} = \frac12 g^{\rho\lambda}\left(\partial_\mu 
g_{\nu\rho} + \partial_\nu g_{\rho\mu} - \partial_\rho g_{\mu\nu}\right)\,.
\end{equation}
In \eqref{distortion}, ${S_{\mu\nu}}^\rho$ is the Cartan torsion tensor,
\begin{equation}
{S_{\mu\nu}}^\rho := {\Gamma^\rho}_{[\mu\nu]}\,, \label{eq:torsdef}
\end{equation}
and $Q_{\lambda\mu\nu}$ is the nonmetricity tensor, defined as
\begin{equation}
Q_{\lambda\mu\nu}:= -\nabla_\lambda g_{\mu\nu} = 
-\partial_\lambda g_{\mu\nu} + {\Gamma^\rho}_{\mu\lambda} g_{\rho\nu} +
{\Gamma^\rho}_{\nu\lambda}g_{\mu\rho} \,.
\end{equation}
Let us also recall the decomposition of the nonmetricity $Q_{\lambda\mu\nu}$ and torsion
${S_{\lambda\mu}}^{\nu}$ in a trace and traceless part. In four dimensions, one has \cite{Iosifidis:2019jgi}
\begin{align}
& Q_{\lambda\mu\nu} = \frac{5}{18} Q_\lambda g_{\mu\nu} - \frac19\tilde{Q}_\lambda g_{\mu\nu} +
\frac49 g_{\lambda(\nu}\tilde{Q}_{\mu)} - \frac19 g_{\lambda(\nu} Q_{\mu)} + \Omega_{\lambda\mu\nu}\,, 
\label{gennm} \\
& {S_{\lambda\mu}}^\nu = \frac{2}{3} {\delta_{[\mu}}^{\nu} S_{\lambda]} + \frac{1}{6} \varepsilon_{\lambda \mu \kappa \rho} g^{\kappa \nu} \tilde{S}^\rho + {Z_{\lambda\mu}}^\nu \,, \label{gentor}
\end{align}
where $Q_\lambda \equiv {Q_{\lambda \mu}}^\mu$ (also known as the Weyl vector) and $\tilde{Q}_\nu := {Q^\mu}_{\mu\nu}$ are the nonmetricity vectors (trace parts of the nonmetricity tensor){,} $S_\lambda :={S_{\lambda \sigma}}^{\sigma}$ is the trace part of the torsion, and $\tilde{S}^\rho$ is the torsion pseudo-vector, while $\Omega_{\lambda\mu\nu}$ denotes the traceless {part} of the nonmetricity and ${Z_{\lambda\mu}}^\nu$ is traceless and such that $Z_{\lambda \mu \nu} = \frac{4}{3} Z_{[\lambda (\mu]\nu)}$ (with $\epsilon^{\lambda \mu \nu \rho} Z_{\lambda \mu \nu}=0$). Here and in the sequel we denote by $\epsilon^{\mu \nu \alpha \beta}$ the Levi-Civita symbol, while $\varepsilon^{\mu \nu \alpha \beta}= \frac{1}{\sqrt{-g}} \epsilon^{\mu \nu \alpha \beta}$ is the Levi-Civita tensor.
In the following, we shall also need the variation of the torsion and nonmetricity w.r.t. the metric and the connection, which is given by \cite{Iosifidis:2019jgi}
\begin{equation}
\delta_g Q_{\rho\alpha\beta} = \partial_\rho\left(g_{\mu\alpha} g_{\nu\beta} \delta g^{\mu\nu}\right) - 
2 g_{\lambda\mu} g_{\nu(\alpha} {\Gamma^\lambda}_{\beta)\rho}\delta g^{\mu\nu}\,, \qquad
\delta_g {S_{\mu\nu}}^\alpha = 0\,, \label{deltag-QT}
\end{equation}
and
\begin{equation}
\delta_\Gamma Q_{\rho\alpha\beta} = 2\delta^\nu_\rho\delta^\mu_{(\alpha} g_{\beta)\lambda}
\delta {\Gamma^\lambda}_{\mu\nu}\,, \qquad
\delta_\Gamma {S_{\alpha\beta}}^\lambda = \delta^{[\mu}_{\alpha}\delta^{\nu]}_{\beta}
\delta {\Gamma^\lambda}_{\mu\nu}\,. \label{deltaconn-QT}
\end{equation}
Finally, we adopt the following definition of the Riemann tensor for a general affine connection ${\Gamma^\lambda}_{\mu \nu}$:
\begin{equation}\label{defRiem}
{R^\mu}_{\nu \alpha \beta} := 2 \partial_{[\alpha} {\Gamma^\mu}_{|\nu|\beta]} + 2 {\Gamma^\mu}_{\rho [\alpha} {\Gamma^\rho}_{|\nu |\beta]} \,.
\end{equation}
Correspondingly, $R_{\mu \nu}={R^\rho}_{\mu \rho \nu}$ and $R=g^{\mu \nu} R_{\mu \nu}$ are the Ricci tensor and the scalar curvature of ${\Gamma^\lambda}_{\mu \nu}$, respectively. We will denote by $\nabla$ the covariant derivative associated with the general affine connection ${\Gamma^\lambda}_{\mu \nu}$, while $\tilde{\nabla}$ will represent the Levi-Civita covariant derivative.

The following decomposition of the Riemann tensor of the general affine connection ${\Gamma^\lambda}_{\mu \nu}$, which holds in four spacetime dimensions, will also be useful in the sequel:
\begin{equation}\label{RiemdecLCTNM}
\begin{split}
R_{\lambda \mu \nu \kappa} & = \frac{1}{2} \left(  g_{\lambda \nu} \tilde{R}_{\mu \kappa} - g_{\lambda \kappa} \tilde{R}_{\mu \nu} - g_{\mu \nu} \tilde{R}_{\lambda \kappa} + g_{\mu \kappa} \tilde{R}_{\lambda \nu} \right) - \frac{1}{6} \tilde{R} \left( g_{\lambda \nu} g_{\mu \kappa} - g_{\lambda \kappa} g_{\mu \nu} \right) + C_{\lambda \mu \nu \kappa} \\
& + \tilde{\nabla}_\kappa N_{\lambda \mu \nu} - \tilde{\nabla}_\nu N_{\lambda \mu \kappa} + N_{\lambda \alpha \nu} {N^\alpha}_{\mu \kappa} - N_{\lambda \alpha \kappa} {N^\alpha}_{\mu \nu} \,,
\end{split}
\end{equation}
where $\tilde{R}_{\mu \nu}$ and $\tilde{R}:=g^{\mu \nu} \tilde{R}_{\mu \nu}$ are the Ricci tensor and Ricci scalar of the Levi-Civita connection $\tilde{\Gamma}^\lambda_{\phantom{\lambda}\mu \nu}$, ${N^\lambda}_{\mu \nu}$ is the distortion tensor given by \eqref{distortion}, and ${C^\lambda}_{\mu \nu \kappa}$ is the so-called Weyl tensor, which fulfills (every tensor contraction between indices of the Weyl tensor gives zero)
\begin{equation}
{C^\lambda}_{\mu \lambda \kappa} = 0 
\end{equation}
and
\begin{equation}
C^{(\lambda \mu) \nu \rho} = 0 \,, \quad C^{\lambda \mu (\nu \rho)} = 0 \,, \quad C^{\lambda \mu \nu \rho} = C^{\nu \rho \lambda \mu} \,, \quad C^{[\lambda \mu \nu \rho]} = 0 \,, \quad g_{\rho \nu} C^{\rho \lambda \mu \nu} = 0 \,.
\end{equation}
The Weyl tensor automatically vanishes in three spacetime dimensions.
We will work in a first order (Palatini) formalism, where the metric $g_{\mu \nu}$ and the connection ${\Gamma^\lambda}_{\mu \nu}$ are treated as independent variables.

Let us also {highlight}, before proceeding, that the following two ways of contracting the torsion with the Levi-Civita tensor are actually not independent:
\beq\label{torscontrNOTindep}
S_{\alpha\beta\mu}\varepsilon^{\alpha\beta\lambda\nu}\,{,} \quad S_{\mu\alpha\beta}\varepsilon^{\alpha\beta\lambda\nu} \,{.}
\eeq 
Indeed, starting with the identity
\beq
\delta^{\nu}_{\mu}\varepsilon^{\alpha\beta\kappa\lambda}+\delta^{\lambda}_{\mu}\varepsilon^{\nu\alpha\beta\kappa}+\delta^{\kappa}_{\mu}\varepsilon^{\lambda\nu\alpha\beta}+\delta^{\beta}_{\mu}\varepsilon^{\kappa\lambda\nu\alpha}+\delta^{\alpha}_{\mu}\varepsilon^{\beta\kappa\lambda\nu}=0
\eeq
and contracting the latter with $S_{\alpha\beta\kappa}$, we easily obtain
\beq\label{idtordual}
-\delta^{\nu}_{\mu}\tilde{S}^{\lambda}+\delta^{\lambda}_{\mu}\tilde{S}^{\nu}+S_{\alpha\beta\mu}\varepsilon^{\alpha\beta\lambda\nu}+2 S_{\mu\alpha\beta}\varepsilon^{\alpha\beta\lambda\nu}=0 \,,
\eeq
indicating their dependence. Note that {substituting} \eqref{gentor} into \eqref{idtordual} the latter {results to be} identically satisfied (it reduces to the trivial identity) and it does not give any constraint on the torsion tensor, as expected.

\section{A parity violating {Metric-Affine} {gravitational} model involving the Holst term}\label{EHHolst}

As a first model we shall consider a {MAG} action consisting of the {EH} term along with the parity violating Holst term. {The first part of this section is a review of the same model studied in \cite{Iosifidis:2019jgi} just in the presence of torsion; here we also consider the presence of nonmetricity and we analyze the complete {Metric-Affine} theory in detail, obtaining new results.} The total action reads
\beq
S_{\text{EH}+\text{Holst}}=\frac{1}{2 \kappa} \int d^{4}x \sqrt{-g}\Big( R+\alpha \varepsilon^{\mu\nu\alpha\beta}R_{\mu\nu\alpha\beta}\Big) \,, \label{S1}
\eeq
where $\kappa=8\pi G$ is the gravitational constant and where $\alpha$ is a dimensionless constant parameter. $R=g^{\mu \nu}R_{\mu \nu}\left(\Gamma \right)$ is the scalar curvature for a general affine connection $\Gamma$ and ${R^\lambda}_{\mu \nu \rho} = {R^\lambda}_{\mu \nu \rho} \left( \Gamma \right)$ is the Riemann tensor whose definition is given in \eqref{defRiem}.

Some comments are now in order. Firstly, note that {\eqref{S1}} is the most general MAG action that is linear in the Riemann tensor. Secondly, both the EH and the additional Holst terms are projective invariant, meaning that, as it can be easily checked, the action \eqref{S1} is invariant under projective transformations\footnote{The latter are defined as those
transformations of the affine connection that leave the autoparallels of vectors invariant up to 
reparametrizations of the affine parameter \cite{Iosifidis:2018zjj}.} of the connection
\begin{equation}\label{projectivetransf}
{\Gamma^\lambda}_{\mu\nu}\mapsto {\Gamma^\lambda}_{\mu\nu} + 
{\delta^\lambda}_\mu\xi_\nu \,,
\end{equation}
where $\xi_\mu$ is an arbitrary one-form.
As a result, the connection field equations will have a vanishing trace in the first two indices. This means that there will be an undetermined {vector} degree of freedom and the connection can be solved only modulo the projective mode. 

Having clarified this{,} let us now proceed with the field equations. Variation w.r.t. the metric gives
\beq
R_{(\mu\nu)}-\frac{1}{2}g_{\mu\nu}R-\alpha \varepsilon_{(\nu}^{\;\;\;\alpha\beta\gamma}R_{\mu)\alpha\beta\gamma}=0 \,, \label{deltag}
\eeq
where the Riemann tensor ${R^\lambda}_{\mu \nu \rho}$ of the affine connection ${\Gamma^\lambda}_{\mu \nu}$ can be written in terms of its Levi-Civita and torsion and nonmetricity as given in \eqref{RiemdecLCTNM}.
By contracting {\eqref{deltag}} with $g_{\mu\nu}$ we get the relation
\beq
\alpha \varepsilon^{\mu\nu\alpha\beta}R_{\mu\nu\alpha\beta}=- R \,. \label{contrdeltag}
\eeq
Note that, using \eqref{contrdeltag}, \eqref{deltag} can also be rewritten as
\begin{equation}\label{deltagNEW}
R_{(\mu\nu)}+\frac{\alpha}{2}g_{\mu\nu} \varepsilon^{\rho \sigma \alpha\beta}R_{\rho \sigma \alpha\beta} -\alpha \varepsilon_{(\nu}^{\;\;\;\alpha\beta\gamma}R_{\mu)\alpha\beta\gamma}=0 \,. 
\end{equation}
On the other hand, varying the action w.r.t the connection we obtain
\beq
P_{\lambda}^{\;\;\;\mu\nu}+2 \alpha \left( {-} \frac{1}{\sqrt{-g}}(2S_{\alpha}-\nabla_{\alpha})(\sqrt{-g}\varepsilon_{\lambda}^{\;\;\mu\alpha\nu})-\varepsilon_{\lambda}^{\;\;\mu\alpha\beta} S_{\alpha\beta}^{\;\;\;\;\nu} \right)=0 \,, \label{Cvar}
\eeq
where
\beq
P_{\lambda}^{\;\;\;\mu\nu}=-\frac{\nabla_{\lambda}(\sqrt{-g}g^{\mu\nu})}{\sqrt{-g}}+\frac{\nabla_{\sigma}(\sqrt{-g}g^{\mu\sigma})\delta^{\nu}_{\lambda}}{\sqrt{-g}} +2(S_{\lambda}g^{\mu\nu}-S^{\mu}\delta_{\lambda}^{\nu}+g^{\mu\sigma}S_{\sigma\lambda}^{\;\;\;\;\nu})
\eeq
is the so-called Palatini tensor, which is traceless in $\mu{,}\lambda$ ({that is} $P_{\mu}^{\;\;\;\mu\nu}=0$). The following formula giving the explicit form of the Palatini tensor will be useful in our calculations (see \cite{Iosifidis:2019jgi} for details on its derivation):
\begin{equation}\label{Palatiniexpl}
\begin{split}
P_{\lambda}^{\;\;\;\mu\nu} & = - {\Omega_\lambda}^{\mu \nu} + \frac{1}{3} g^{\mu \nu} \left( \frac{2}{3} Q_\lambda + \frac{1}{3} \tilde{Q}_\lambda + 4 S_\lambda \right) + \frac{1}{9} {\delta_\lambda}^\nu \left(-4 Q^\mu + 7 \tilde{Q}^\mu \right) + \frac{1}{9} {\delta_\lambda}^\mu  \left( \frac{1}{2} Q^\mu - 2 \tilde{Q}^\nu \right) \\
& - \frac{1}{3} {\varepsilon_\lambda}^{\mu \nu \rho} \tilde{S}_\rho - 2 {Z_\lambda}^{\mu \nu} \,.
\end{split}
\end{equation}
{The} additional contribution from the Holst term in \eqref{Cvar} also has a vanishing trace in $\mu,\lambda${,} as we already mentioned. In fact, there is a somewhat more convenient way to derive the connection field equations in another form. Indeed, using the identity\footnote{See \cite{Iosifidis:2019jgi}; here we also correct a misprint appearing there.}
\beq
\varepsilon^{\mu\nu\alpha\beta}R_{\mu\nu\alpha\beta}=2 \tilde{\nabla}_{\alpha}\tilde{S}^{\alpha}+2 \varepsilon^{\mu\nu\alpha\beta}S_{\alpha\beta}^{\;\;\;\;\lambda}( Q_{\mu\nu\lambda}+S_{\mu\nu\lambda}) \,,
\eeq
disregarding surface terms we arrive at
\beq
P_{\lambda}^{\;\;\;\mu\nu}+2 \alpha \Big( \varepsilon^{\mu\nu\alpha\beta}Q_{\alpha\beta\lambda}+ \varepsilon^{\mu\nu\alpha\beta}S_{\alpha\beta\lambda}-\varepsilon_{\lambda}^{\;\;\nu\alpha\beta}S_{\alpha\beta}^{\;\;\;\;\mu}\Big)=0 \,. \label{varconnNEW}
\eeq
Let us observe that \eqref{Cvar} and \eqref{varconnNEW} actually coincide, due to the following identity which holds in four dimensions{:}
\begin{equation}
Z_{\lambda \mu \nu} = \varepsilon_{\lambda \mu \rho \sigma} \tilde{Z}^{\rho \sigma}_{\phantom{\rho \sigma} \nu} \,,
\end{equation}
where $\tilde{Z}_{\lambda \mu}^{\phantom{\lambda \mu} \nu}$ is a tensor such that
\begin{equation}
\tilde{Z}_{[\lambda \mu \nu]} = 0 \,.
\end{equation}
Now, contracting \eqref{varconnNEW} one time in $\lambda,\nu$ and another with $g_{\mu\nu}$, after some relabeling, we obtain
\beq\label{tr1}
P^{\mu}+2\alpha \tilde{S}^{\mu}=0
\eeq
and
\beq\label{tr2}
\tilde{P}^{\mu}-2\alpha \tilde{S}^{\mu}=0 \,,
\eeq
where 
\begin{equation}\label{Ptr1}
P^{\mu}\equiv P_{\nu}^{\;\;\;\mu\nu}=3\left( \tilde{Q}^{\mu}-\frac{1}{2}Q^{\mu}\right) -4 S^{\mu}
\end{equation}
and
\begin{equation}\label{Ptr2}
\tilde{P}_{\lambda} \equiv g_{\mu\nu}P_{\lambda}^{\;\;\;\mu\nu}=\frac{1}{2}Q_{\lambda}+\tilde{Q}_{\lambda}+4S_{\lambda} \,.
\end{equation}
Notice that the traces coming from the variation of the Holst term are the same up to a minus sign. To be more specific, defining
\begin{gather}
H_{\lambda}^{\;\;\mu\nu}:=\frac{ \alpha}{\sqrt{-g}}\frac{\delta (\sqrt{-g}\varepsilon^{\mu\nu\alpha\beta}R_{\mu\nu\alpha\beta}) }{\delta \Gamma^{\lambda}_{\;\;\;\;\mu\nu}}
=2 \alpha \Big( \varepsilon^{\mu\nu\alpha\beta}Q_{\alpha\beta\lambda}+ \varepsilon^{\mu\nu\alpha\beta}S_{\alpha\beta\lambda}-\varepsilon_{\lambda}^{\;\;\nu\alpha\beta}S_{\alpha\beta}^{\;\;\;\;\mu}\Big) \,,
\end{gather}
we observe that
\beq
H_{\lambda}^{\;\;\mu\lambda}=-H^{\mu\alpha\beta}g_{\alpha\beta}=2 \alpha \tilde{S}^{\mu} \,.
\eeq
That is, the two contractions are linearly dependent. This is hardly a coincidence and implies that a larger symmetry is at play here. Indeed, using the results of \cite{Iosifidis:2019fsh} the above relation between the traces implies that the Holst term is invariant under the connection transformation
\beq
\Gamma^{\lambda}_{\;\;\mu\nu}\mapsto \Gamma^{\lambda}_{\;\;\mu\nu}+\delta^{\lambda}_{\nu}\zeta_{\mu}+\zeta^{\lambda}g_{\mu\nu} \,,
\eeq
where $\zeta_{\mu}$ is an arbitrary one-form vector.
Using the terminology of \cite{Iosifidis:2019fsh}, this is a constrained vectorial transformation of the connection. Recalling that the Holst term is also invariant under projective transformations, we conclude that the latter is invariant under the general class of transformations
\beq
\Gamma^{\lambda}_{\;\;\mu\nu}\mapsto \Gamma^{\lambda}_{\;\;\mu\nu}+\delta^{\lambda}_{\mu}\xi_{\nu}+\delta^{\lambda}_{\nu}\zeta_{\mu}+\zeta^{\lambda}g_{\mu\nu} \,, \label{genclasstr}
\eeq
where $\xi_{\mu}$ and $\zeta_{\mu}$ are arbitrary independent {one-forms}. Notice, however, that the total action \eqref{S1} is not invariant under \eqref{genclasstr}.

We can now move on to the analysis of the field equations.
In the following analysis, we shall discard the trivial case $\alpha=0$.\footnote{{The trivial case $\alpha=0$} would correspond to the absence of the {parity} violating Holst term in the action and would lead to a purely Levi-Civita \eqref{lcconn} connection by choosing an appropriate gauge parameter for the projective mode.}
Combining  \eqref{tr1}-\eqref{Ptr2}, we obtain
\begin{equation}\label{Q}
Q_\lambda = \frac{8}{3} \alpha \tilde{S}_\lambda - \frac{16}{3} S_\lambda 
\end{equation}
and
\begin{equation}\label{Qt}
\tilde{Q}_\lambda = \frac{1}{4} Q_\lambda = \frac{2}{3} \alpha \tilde{S}_\lambda - \frac{4}{3} S_\lambda  \,.
\end{equation}
Using all the above, together with the decomposition of the nonmetricity and torsion tensors, namely \eqref{gennm} and \eqref{gentor}, eq. \eqref{varconnNEW} becomes
\begin{equation}\label{intermstep}
2 \alpha {\varepsilon^{\mu \nu}}_{\rho \sigma} {\Omega^{\rho \sigma}}_\lambda - {\Omega_\lambda}^{\mu \nu} - \frac{1}{3} \left( 1 + 4 \alpha^2 \right) {\varepsilon_\lambda}^{\mu \nu \rho} \tilde{S}_\rho + 2 \alpha \left( \varepsilon^{\mu \nu \rho \sigma} Z_{\rho \sigma \lambda} - {\varepsilon_\lambda}^{\nu \rho \sigma} {Z_{\rho \sigma}}^\mu \right) - 2 {Z_\lambda}^{\mu \nu} = 0 \,.
\end{equation}
After some algebraic manipulation and applying contractions (involving also contractions with the Levi-Civita tensor) on \eqref{intermstep}, we find
\begin{equation}\label{St}
\left( 1 + 4 \alpha^2 \right) \tilde{S}_\mu = 0 \,,
\end{equation}
\begin{equation}\label{Omega}
\Omega_{\lambda \mu \nu} = - 2 \left(1 + 4 \alpha^2 \right) Z_{\lambda (\mu \nu)} \,,
\end{equation}
and
\begin{equation}\label{Zeta}
{Z_{\mu \nu}}^\lambda = \frac{1+12 \alpha^2}{4 \alpha} \varepsilon_{\mu \nu \rho \sigma} Z^{\lambda \rho \sigma} \,.
\end{equation}
Let us observe that using \eqref{Q}-\eqref{Zeta}, we have that \eqref{contrdeltag} yields 
\begin{equation}\label{RLCzero}
\tilde{R} = 0 \,,
\end{equation}
where the vanishing scalar curvature $\tilde{R}$ is the Ricci scalar of the Levi-Civita connection, and that \eqref{intermstep} is identically satisfied. Then, summarizing, we are left with the above equations together with \eqref{deltag}. Moreover, we can write the provisional form of the connection as follows ({substituting} \eqref{gennm} and \eqref{gentor} into \eqref{distortion}, using \eqref{Q} and \eqref{Qt}, and plugging the result into \eqref{gendecompaffconn}): 
\begin{equation}\label{connformprov}
\begin{split}
{\Gamma^\lambda}_{\mu \nu} & = \tilde{\Gamma}^\lambda_{\phantom{\lambda} \mu \nu} + \frac{1}{6} g^{\lambda \kappa}\varepsilon_{\kappa \mu \nu \rho} \tilde{S}^\rho - \frac{1}{3} \alpha g_{\mu \nu}\tilde{S}^\lambda + \frac{2}{3} \alpha {\delta^\lambda}_{(\mu} \tilde{S}_{\nu)} - \frac{2}{3} {\delta^\lambda}_\mu S_\nu \\
&  - \frac{1}{2} {\Omega^\lambda}_{\mu \nu} + {\Omega_{(\mu \nu)}}^\lambda - 2 {Z^\lambda}_{\mu \nu}  \,,
\end{split}
\end{equation}
where \eqref{Omega} and \eqref{Zeta} hold and where $\tilde{\Gamma}^\lambda_{\phantom{\lambda} \mu \nu}$ is the Levi-Civita connection defined in \eqref{lcconn}.

Now, we can see that, if we consider $1+12 \alpha^2 \neq 0$ (that is, $\alpha \neq \pm \frac{{i}}{2\sqrt{3}}$), we can contract \eqref{Zeta} with the Levi Civita tensor, obtaining
\begin{equation}\label{Zetadual}
{Z_{\alpha \beta}}^\lambda = \frac{2 \alpha}{1+12 \alpha^2} \varepsilon_{\alpha \beta \mu \nu} Z^{\mu \nu \lambda} \,.
\end{equation}
Then, comparing \eqref{Zeta} with \eqref{Zetadual} and using the fact that $Z_{[\mu \nu \lambda]}=0 \, \Rightarrow \, Z_{\mu \nu \lambda} = -2 Z_{\lambda [\mu \nu]}$, we get the following equation:
\begin{equation}
\left( \frac{1+12 \alpha^2}{4 \alpha} \right)^2 = -1 \,,
\end{equation}
whose solution {reads}
\begin{equation}\label{solAnotzero}
\alpha = \pm \frac{{i}}{2} \quad \vee \quad \alpha = \pm \frac{{i}}{6} \,.
\end{equation}
In any other case, one has $Z_{\lambda \mu \nu}=0$.
Let us now analyze the different solutions.

\subsection{Case $\alpha = \pm \frac{{i}}{2\sqrt{3}}$}

In this case, \eqref{Omega} and \eqref{Zeta} become, respectively
\begin{equation}\label{Omegazero}
\Omega_{\lambda \mu \nu} = 0  \,,
\end{equation}
\begin{equation}\label{Zetazero}
Z_{\lambda \mu \nu} = 0 \,.
\end{equation}
Moreover, from \eqref{St} we get
\begin{equation}\label{Stzero}
\tilde{S}_\lambda = 0 \,,
\end{equation}
which, plugged into \eqref{Q} and \eqref{Qt}, implies
\begin{equation}\label{QS}
Q_\lambda = - \frac{16}{3} S_\lambda 
\end{equation}
and
\begin{equation}\label{QtS}
\tilde{Q}_\lambda = - \frac{4}{3} S_\lambda \,.
\end{equation}
Thus, we are left with a single independent vector, $S_\lambda$, and, using \eqref{Omegazero}-\eqref{QtS}, we find that the final form of the connection \eqref{connformprov} becomes
\begin{equation}\label{finconn1}
{\Gamma^\lambda}_{\mu \nu} = \tilde{\Gamma}^\lambda_{\phantom{\lambda} \mu \nu} - \frac{2}{3} {\delta^\lambda}_\mu S_\nu \,.
\end{equation}
Let us now observe that the torsion vector $S_\lambda$ in \eqref{finconn1} can be gauged away through a projective transformation of the connection \eqref{projectivetransf} with $\xi_\mu = \frac{2}{3} S_\mu$, that is
\begin{equation}\label{prtrtor}
{\Gamma^\lambda}_{\mu\nu}\mapsto {\Gamma^\lambda}_{\mu\nu} +  \frac{2}{3} {\delta^\lambda}_\mu S_\nu \,.
\end{equation}
This means that, exploiting \eqref{prtrtor}, the connection of the theory reduces to the Levi-Civita connection \eqref{lcconn}, namely
\begin{equation}\label{finconn1LC}
{\Gamma^\lambda}_{\mu \nu} = \tilde{\Gamma}^\lambda_{\phantom{\lambda} \mu \nu}  \,.
\end{equation}
Notice that, in this case, using the above equations together with \eqref{RLCzero} we also find that \eqref{deltag} becomes
\begin{equation}\label{RLCmnzero}
\tilde{R}_{\mu \nu} = 0 \,,
\end{equation}
where $\tilde{R}_{\mu \nu}$ is the Ricci tensor of the Levi Civita connection.
{Therefore, in this case the theory results to be equivalent to vacuum GR.}

\subsection{Case $\alpha = \pm \frac{{i}}{2}$} 

{Let us mention, before proceeding, that} these particular values of $\alpha$ are relevant in the context of the canonical 3+1 Hamiltonian formulation of the Holst action, where, in particular, $\alpha=-\frac{{i}}{2}$ happens to correspond to Ashtekar variables, formulating (complex) GR as a special type of Yang-Mills gauge theory. The action was seen to correspond to the Palatini action with the curvature tensor replaced by its self-dual part only \cite{Samuel:1987td,Jacobson:1987yw,Jacobson:1988yy,Goldberg:1988ir,Henneaux:1989hf}.
{In the present framework, in this case we have} that \eqref{St} is identically satisfied and \eqref{Q}, \eqref{Qt}, and \eqref{Zeta} respectively become
\begin{equation}\label{QStS}
Q_\lambda = \pm \frac{4 {i} }{3} \tilde{S}_\lambda - \frac{16}{3} S_\lambda \,,
\end{equation}
\begin{equation}\label{QtStS}
\tilde{Q}_\lambda = \pm \frac{{i} }{3} \tilde{S}_\lambda - \frac{4}{3} S_\lambda \,,
\end{equation}
\begin{equation}\label{Zeta2}
Z_{\mu \nu \lambda} = \pm {i} \varepsilon_{\mu \nu \rho \sigma} {Z_\lambda}^{\rho \sigma} \,.
\end{equation}
Moreover, eq. \eqref{Omega} reduces to \eqref{Omegazero}.
In particular, we can see that we are left with two independent vectors, say $\tilde{S}_\lambda$ and $S_\lambda$. Using \eqref{Omegazero}, \eqref{QStS}, and \eqref{QtStS}, we find that the final form of the connection \eqref{connformprov} now becomes
\begin{equation}\label{finconn2}
{\Gamma^\lambda}_{\mu \nu} = \tilde{\Gamma}^\lambda_{\phantom{\lambda} \mu \nu} + \frac{1}{6} {\varepsilon^\lambda}_{\mu \nu \rho} \tilde{S}^\rho \pm \frac{{i}}{6} \left( 2 {\delta^\lambda}_{(\mu} \tilde{S}_{\nu)} - g_{\mu \nu}\tilde{S}^\lambda \right) - \frac{2}{3} {\delta^\lambda}_\mu S_\nu  - 2 {Z^\lambda}_{\mu \nu} \,,
\end{equation}
with $Z_{\lambda \mu \nu}$ obeying \eqref{Zeta2}. This form for the connection seems to be rather complicated but, as we will see in the sequel, exploiting {a remarkable} symmetry of the action one can remove the torsion contribution, ending up with a final connection that is the purely Levi-Civita one.

\subsubsection{The enlarged symmetry corresponding to the case $\alpha= \pm \frac{i}{2}$}

The peculiar cases $\alpha=\pm \frac{i}{2}$ are very interesting for many reasons. Some of them we outlined in the {I}ntroduction. Another one is the following{:} As we have shown, for $\alpha=\pm \frac{i}{2}$ the torsion pseudo-vector $\tilde{S}^{\mu}$ is left completely unspecified. This is {not a} coincidence and there is a deeper reason why this is so. As we will show {below,} this very reason is again some symmetry obeyed by the total action \eqref{S1}. To start with, let us first define the total off-shell variation
\beq
W_{\lambda}^{\;\;\mu\nu}:=\frac{\delta}{\delta \Gamma^{\lambda}_{\;\;\;\mu\nu}}\Big(R+\alpha \varepsilon^{\mu\nu\alpha\beta}R_{\mu\nu\alpha\beta}\Big) \,,
\eeq
where, at this point, {we are not making any assumption on} the value of $\alpha$. Defining the traces \cite{Iosifidis:2019fsh}
\beq
{W_{(1)}^{\mu}}:=W_{\lambda}^{\;\;\lambda\mu} \;, \;\; W_{(2)}^{\mu}:=W_{\lambda}^{\;\;\mu\lambda}\;, \;\; W^{\mu}_{(3)}:=W^{\mu\alpha\beta}g_{\alpha\beta} \,,
\eeq
we immediately have that ${W_{(1)}^{\mu}}\equiv 0$ since both the {EH} and the Holst {terms} are independently projective invariant. {Regarding} the other two traces, by subtracting them we find
\beq
W_{(2)}^{\mu}-W_{(3)}^{\mu}=2 (\tilde{Q}^{\mu}-Q^{\mu}-4S^{\mu})+4 \alpha \tilde{S}^{\mu} \,. \label{t1}
\eeq
Moreover, we may also compute the contraction of $W_{\alpha\mu\nu}$ with the Levi-Civita tensor, which gives us the pseudo-vector 
\beq
\varepsilon^{\alpha\beta\gamma\mu}W_{\alpha\beta\gamma}=  2 \tilde{S}^{\mu}- 4 \alpha (\tilde{Q}^{\mu}-Q^{\mu}-4S^{\mu})  \,. \label{t2}
\eeq
Notice now that for generic values of {$\alpha$} the traces \eqref{t1} and \eqref{t2} are linearly independent unless $\alpha=\pm \frac{i}{2}$, in which case they become linearly dependent and it holds that
\beq
W_{(2)}^{\mu}-W_{(3)}^{\mu}\mp i \varepsilon^{\alpha\beta\gamma\mu}W_{\alpha\beta\gamma}=0 \,.
\eeq
Then, using {the} Theorem of \cite{Iosifidis:2019fsh}, we can conclude that in this case the total action \eqref{S1} is also invariant under connection transformations of the form
\beq
\Gamma^{\lambda}_{\;\;\;\mu\nu} \mapsto \Gamma^{\lambda}_{\;\;\;\mu\nu}+\delta^{\lambda}_{\nu}\psi_{\mu}-\psi^{\lambda}g_{\mu\nu}\mp {i} \varepsilon^{\lambda}_{\;\;\mu\nu\alpha}\psi^{\alpha} \,, \label{gentronlypsi}
\eeq
where $\psi_{\mu}$ is an arbitrary one-form vector. It is worth stressing out that this symmetry is another gauge symmetry for the total action \eqref{S1} on top (and independent) of the projective invariance. Collecting the above results{,} we are led to the following conclusion:
\begin{prop}
	The scalar combination
	\beq
	R \pm \frac{{i}}{2} \varepsilon^{\mu\nu\alpha\beta}R_{\mu\nu\alpha\beta} \label{RH}
	\eeq
	is invariant under the two-parameter gauge transformations of the connection
	\beq
	\Gamma^{\lambda}_{\;\;\;\mu\nu} \mapsto \Gamma^{\lambda}_{\;\;\;\mu\nu} +\delta^{\lambda}_{\mu}\xi_{\nu} +\delta^{\lambda}_{\nu}\psi_{\mu}-\psi^{\lambda}g_{\mu\nu}\mp {i} \varepsilon^{\lambda}_{\;\;\mu\nu\alpha}\psi^{\alpha} \,, \label{ensym}
	\eeq
 where $\xi_{\mu}$ and $\psi_{\mu}$ are arbitrary independent one-form vectors. That is, for $\alpha =\pm \frac{i}{2}$ the total MAG action \eqref{S1} consisting of the {EH contribution plus} the Holst term admits the enlarged symmetry \eqref{ensym} generalizing projective invariance.
 \begin{proof}
 	{Both the EH and the Holst terms} are independently projective invariant. {Consequently,} any combination of {them} will also be so. That is, \eqref{RH} is invariant under 
 	\beq
 	\Gamma^{\lambda}_{\;\;\;\mu\nu} \mapsto \Gamma^{\lambda}_{\;\;\;\mu\nu}+\delta^{\lambda}_{\mu}\xi_{\nu} \,.
 	\eeq
 	In addition, as we {have shown above,} the combination in \eqref{RH}, but not each term separately, is invariant under
 	\beq
 	\Gamma^{\lambda}_{\;\;\;\mu\nu} \mapsto \Gamma^{\lambda}_{\;\;\;\mu\nu}+\delta^{\lambda}_{\nu}\psi_{\mu}-\psi^{\lambda}g_{\mu\nu}\mp {i} \varepsilon^{\lambda}_{\;\;\mu\nu\alpha}\psi^{\alpha} \,. \label{g2}
 	\eeq
 	Then, combining the above two results{,} we conclude that the combination \eqref{RH} is invariant under the general class of transformations {\eqref{ensym}.}
 \end{proof}
\end{prop}
The last result has dramatic consequences when applied to \eqref{finconn2}. Indeed, after some trivial rearrangement, \eqref{finconn2} reads
\beq
{\Gamma^\lambda}_{\mu \nu} = \tilde{\Gamma}^\lambda_{\phantom{\lambda} \mu \nu} \pm \frac{i}{6} \left(\delta_{\nu}^{\lambda}\tilde{S}_{\mu}- g_{\mu \nu}\tilde{S}^\lambda \mp {i} {\varepsilon^\lambda}_{\mu \nu \rho} \tilde{S}^\rho   \right) + \frac{1}{3}\left( -2S_\nu \pm \frac{i}{2}\tilde{S}_{\nu} \right) {\delta^\lambda}_\mu   - 2 {Z^\lambda}_{\mu \nu} \,.
\eeq
Note now that the second parenthesis in the right-hand side of the above equation can be gauged away by using the projective freedom and appropriately picking $\xi_{\mu}$. In addition, the first parenthesis can also be gauged away by using the extended invariance of the action under \eqref{g2} for $\alpha= \pm \frac{i}{2}$. Fixing these two gauges, the connection takes the simpler form
\beq 
{\Gamma^\lambda}_{\mu \nu} = \tilde{\Gamma}^\lambda_{\phantom{\lambda} \mu \nu} - 2 {Z^\lambda}_{\mu \nu} \,, \label{connonlytoraftervar}
\eeq
where the gauge-fixing has enabled us to remove all the vectorial parts. 
Finally, one can prove that the action \eqref{S1} with the choice $\alpha=\pm \frac{{i}}{2}$ is also invariant under the following transformation of the connection\footnote{One can prove this by noting that $W_{\lambda \mu \nu}=\pm {i} \varepsilon_{\lambda \mu \rho \sigma} {W_\nu}^{\rho \sigma}$ and $W_{[\lambda \mu \nu]} =0$. Then{,} using the results of \cite{Iosifidis:2019fsh} we obtain the invariance under the associated connection transformation as given by (\ref{trreabsZ}).}
\begin{equation}\label{trreabsZ}
{\Gamma^\lambda}_{\mu \nu} \mapsto {\Gamma^\lambda}_{\mu \nu} + {V^\lambda}_{\mu \nu} \,,
\end{equation}
where $V_{\lambda \mu \nu} $ is a tensor such that
\begin{equation}\label{Vtensprop}
V_{\lambda \mu \nu} = \pm {i} \varepsilon_{\lambda \mu \rho \sigma} {V_\nu}^{\rho \sigma} \,, \quad V_{[\lambda \mu \nu]} =0 \,.
\end{equation}
Then, exploiting \eqref{trreabsZ} with the choice 
\begin{equation}\label{VZ}
{V^\lambda}_{\mu \nu} = 2 {Z^\lambda}_{\mu \nu} \,,
\end{equation}
the final connection simply reduces to the purely Levi-Civita one, \eqref{lcconn}, meaning that we are left with \eqref{finconn1LC}. In other words, we have found that when $\alpha=\pm \frac{{i}}{2}$ also {the} $Z_{\lambda \mu \nu}$ tensor is non-dynamical and can be reabsorbed (consistently set to zero), meaning that eq. \eqref{deltag} reduces to \eqref{RLCmnzero} {and that the theory results, again, to be equivalent to GR in vacuum.}
{Let us conclude by highlighting the following result we have obtained so far:}
{\begin{prop}
	The scalar combination \eqref{RH} is invariant under the gauge transformations of the connection
	\begin{equation}\label{enlsymm}
\Gamma^{\lambda}_{\;\;\;\mu\nu} \mapsto \Gamma^{\lambda}_{\;\;\;\mu\nu}+\delta^{\lambda}_{\nu}\psi_{\mu}-\psi^{\lambda}g_{\mu\nu}\mp {i} \varepsilon^{\lambda}_{\;\;\mu\nu\alpha}\psi^{\alpha}+\delta^{\lambda}_{\mu}\xi_{\nu} + {V^\lambda}_{\mu \nu} \,,
\end{equation}
where ${V^\lambda}_{\mu \nu}$ is a tensor fulfilling \eqref{Vtensprop} and where $\psi_\mu$ and $\xi_\mu$ are arbitrary one-form vectors. That is, for $\alpha =\pm \frac{i}{2}$ the total MAG action \eqref{S1} consisting of the {EH contribution plus} the Holst term admits the enlarged symmetry \eqref{enlsymm}, which generalize projective invariance. The above defines an  equivalence class of connections with $4+4+16=24$ elements.
 \begin{proof}
 	As we have previously proved, for $\alpha =\pm \frac{i}{2}$ the total MAG action \eqref{S1} is invariant under \eqref{ensym}. Combining this with the above discussion on the transformations involving the tensor ${V^\lambda}_{\mu \nu}$, we can conclude that the combination \eqref{RH} is invariant under the general class of transformations \eqref{enlsymm}.
 \end{proof}
\end{prop}}

\subsection{Case $\alpha = \pm \frac{{i}}{6}$} 

Here, from \eqref{St} we get \eqref{Stzero}, while \eqref{Q} and \eqref{Qt} respectively reduce to \eqref{QS} and \eqref{QtS}. Moreover, \eqref{Zeta} becomes
\begin{equation}\label{Zeta3}
Z_{\mu \nu \lambda} = \mp {i} \varepsilon_{\mu \nu \rho \sigma} {Z_\lambda}^{\rho \sigma} \,,
\end{equation}
and \eqref{Omega} gives
\begin{equation}\label{Omega3}
\Omega_{\lambda \mu \nu} = - \frac{8}{3} Z_{\lambda (\mu \nu)} \,.
\end{equation}
Thus, in this case, the final form of the connection \eqref{connformprov} reads
\begin{equation}\label{finconn3}
{\Gamma^\lambda}_{\mu \nu} = \tilde{\Gamma}^\lambda_{\phantom{\lambda} \mu \nu} - \frac{2}{3} {\delta^\lambda}_\mu S_\nu  + \frac{2}{3} {Z^\lambda}_{\nu \mu} + \frac{2}{3} {Z_{\mu \nu}}^\lambda \,.
\end{equation}
As in the first case, here we have just one independent vector, $S_\lambda$, and we can thus exploit the projective transformation \eqref{prtrtor} to reabsorb it, obtaining
\begin{equation}\label{finconn3fin}
{\Gamma^\lambda}_{\mu \nu} = \tilde{\Gamma}^\lambda_{\phantom{\lambda} \mu \nu} + \frac{2}{3} {Z^\lambda}_{\nu \mu} + \frac{2}{3} {Z_{\mu \nu}}^\lambda \,.
\end{equation}
Finally, using the above equations, together with \eqref{RLCzero}, eq. \eqref{deltag} yields
\begin{equation}\label{deltagFin3}
\tilde{R}_{\mu \nu} + \frac{1}{3} \left( - \frac{1}{2} g_{\mu \nu} Z_{\alpha \beta \gamma} Z^{\alpha \beta \gamma} + 2 {Z_{\mu}}^{\alpha \beta} Z_{\nu \alpha \beta} \right) =0  \,,
\end{equation}
where $Z_{\lambda \mu \nu}$ obeys \eqref{Zeta3}.

Now, let us observe that one can prove that the action \eqref{S1} with the choice $\alpha=\pm \frac{{i}}{6}$ results to be invariant under the following transformation of the connection
\begin{equation}\label{trreabsZ2}
{\Gamma^\lambda}_{\mu \nu} \mapsto {\Gamma^\lambda}_{\mu \nu} + {V^\lambda}_{\nu \nu} + {V_{\mu \nu}}^\lambda \,,
\end{equation}
where $V_{\lambda \mu \nu} $ is a tensor such that
\begin{equation}
V_{\lambda \mu \nu} = \mp {i} \varepsilon_{\lambda \mu \rho \sigma} {V_\nu}^{\rho \sigma} \,, \quad V_{[\lambda \mu \nu]} =0 \,.
\end{equation}
Then, exploiting \eqref{trreabsZ2} and performing the choice 
\begin{equation}\label{VZ2}
{V^\lambda}_{\mu \nu} = - \frac{2}{3} {Z^\lambda}_{\mu \nu} \,,
\end{equation}
the final connection simply reduces to the purely Levi-Civita one, that is we are left once again with \eqref{finconn1LC}. Consequently, \eqref{deltagFin3} can be consistently reduced to \eqref{RLCmnzero} {and the theory is equivalent to vacuum GR}.

\subsection{Cases $\alpha \neq \pm\frac{i}{2} \wedge \alpha \neq \pm\frac{i}{6}$}

In all these remaining cases (which actually include also the case $\alpha = \pm \frac{i}{2\sqrt{3}}$), as we have previously proved, we are left with \eqref{Zetazero} and, thus, also with \eqref{Omegazero}. Moreover, we have \eqref{Stzero}, which also implies \eqref{QS} and \eqref{QtS}. Finally, the remaining independent vector $S_\lambda$ can be eliminated by exploiting projective invariance, as we have already discussed above, and we end up with a connection that is purely Levi-Civita, together with eqs. \eqref{RLCzero} and \eqref{RLCmnzero}.

\paragraph{Summary of the results for the {Metric-Affine} EH + Holst theory.} We have shown that for $\alpha=\pm \frac{i}{2}$, interestingly, the action \eqref{S1} {is invariant under connection transformations of the form \eqref{enlsymm}, which, in particular, generalize projective invariance.} This peculiar feature is reported here for the first time and it allows to reduce the final form of the connection, by appropriate gauge choices, to the purely Levi-Civita one. Additionally, we have proved that in all the solutions to the theory given by the action \eqref{S1} the final form of the connection can be recast into the form \eqref{lcconn}, that is it can be reduced to the Levi-Civita connection, and the equations coming from the variation w.r.t. the metric reduce to \eqref{RLCzero} and \eqref{RLCmnzero}. Thus, the theory turns out to be equivalent to GR in vacuum.

\subsection{Comments on the torsionful metric case}

Let us first of all mention, here, that if we set the nonmetricity tensor $Q_{\lambda \mu \nu}$ to zero in the model under analysis after variation of the action \eqref{S1}, we get either vanishing torsion, which leads to a final connection that is the purely Levi-Civita and to \eqref{RLCzero} and \eqref{RLCmnzero}, or a model with vanishing torsion traces 
\begin{equation}
S_\lambda = 0 \,, \quad \tilde{S}_\lambda = 0  \,,
\end{equation}
but non-vanishing traceless part of the torsion. Indeed, considering \eqref{Omega} and setting $\Omega_{\lambda \mu \nu} = 0$, we have that either $Z_{\lambda \mu \nu}=0$ or $\alpha = \pm \frac{{i}}{2}$. In the latter case, $Z_{\lambda \mu \nu}$ obeys \eqref{Zeta2} and the final form of the connection coincides with \eqref{connonlytoraftervar}. Nevertheless, as we have already mentioned, in the case $\alpha = \pm \frac{{i}}{2}$ we can exploit the transformation \eqref{trreabsZ} with the choice \eqref{VZ} to reabsorb the traceless part of the torsion, ending up with a purely Levi-Civita connection \eqref{finconn1LC} also in this case.
Also, we find \eqref{RLCzero} and \eqref{deltag} reduces to \eqref{RLCmnzero}. {Thus, in each of these cases, the final form of the connection is given by the Levi-Civita one and the theory is equivalent to vacuum GR.}

If we now focus on the parity violating model \eqref{S1} by considering a torsionful metric connection from the very beginning (that is to say, directly putting the nonmetricity to zero in \eqref{S1} and restricting ourselves to the so-called Einstein-Cartan framework), what we get is again the purely Levi-Civita connection for the final form of the connection. Indeed, by performing an analysis on the same line of the one previously done, recalling that now
\begin{equation}\label{nonmetzero}
Q_{\lambda \mu \nu}=0
\end{equation}
from the very beginning, one can prove that the equation obtained by varying \eqref{S1} w.r.t. the connection yields \eqref{Stzero} and, consequently, as one can easily show,
\begin{equation}\label{Szero}
S_\lambda = 0 \,,
\end{equation}
together with \eqref{Zetazero}. Then, the final form of the connection results to be \eqref{finconn1LC} and, again, the equations obtained from the variation w.r.t. the metric reduce to \eqref{RLCzero} and \eqref{RLCmnzero}, {that is the theory is equivalent to GR in vacuum.}

\subsection{Observations on the purely {Metric-Affine} Holst action}

As we have already noticed, the {Metric-Affine} Holst term in \eqref{S1} is invariant not only under projective transformations but also under the general class of transformations \eqref{genclasstr}. Moreover, one can prove that it is also invariant under the following transformation of the connection:
\begin{equation}\label{trconnsymm}
{\Gamma^\lambda}_{\mu \nu} \mapsto {\Gamma^\lambda}_{\mu \nu} + {U^\lambda}_{\mu \nu} \,,
\end{equation}
where ${U^\lambda}_{\mu \nu}$ is a completely symmetric tensor. This will be useful in the sequel, where we will study the case in which only the {Metric-Affine} Holst term contributes to the theory.

Thus, let us now consider the action
\beq
S_{\text{Holst}}=\frac{1}{2 \kappa} \int d^{4}x \sqrt{-g} \alpha \varepsilon^{\mu\nu\alpha\beta}R_{\mu\nu\alpha\beta} \,. \label{S12}
\eeq
As usual, we shall always discard the trivial case $\alpha = 0$.
Variation w.r.t. $g^{\mu \nu}$ gives
\beq
\varepsilon_{(\nu}^{\;\;\;\alpha\beta\gamma}R_{\mu)\alpha\beta\gamma}=0 \,. \label{deltag12}
\eeq
By contracting this with $g_{\mu\nu}$, we obtain
\beq
\varepsilon^{\mu\nu\alpha\beta}R_{\mu\nu\alpha\beta}= 0 \,. \label{contrdeltag12}
\eeq
On the other hand, the variation w.r.t. ${\Gamma^\lambda}_{\mu \nu}$ yields
\beq
\varepsilon^{\mu\nu\alpha\beta}Q_{\alpha\beta\lambda}+ \varepsilon^{\mu\nu\alpha\beta}S_{\alpha\beta\lambda}-\varepsilon_{\lambda}^{\;\;\nu\alpha\beta}S_{\alpha\beta}^{\;\;\;\;\mu} =0 \,. \label{varconnNEW12}
\eeq
As we have already mentioned (and can be easily proved), \eqref{varconnNEW12} has a vanishing trace in $\mu,\lambda$. Now, contracting \eqref{varconnNEW12} in $\lambda,\nu$ we get \eqref{Stzero}, while the contraction with $g_{\mu\nu}$ identically vanishes. {Substituting} \eqref{Stzero} into \eqref{varconnNEW12} and contracting the resulting equation with the Levi-Civita tensor ${\varepsilon^\lambda}_{\mu \nu \rho}$, we find
\begin{equation}\label{epscontr12}
\tilde{Q}_\rho = Q_\rho + 4  S_\rho \,,
\end{equation}
which means, together with \eqref{Stzero}, that we are left with two independent vectors, $Q_\rho$ and $S_\rho$.
Plugging \eqref{epscontr12} into the resulting equation mentioned above, after some contractions with the Levi-Civita tensor and making some algebraic manipulation, we end up with \eqref{Zetazero}, together with
\begin{equation}\label{Omcomplsymm}
\Omega_{\lambda \mu \nu} = \Omega_{(\lambda \mu \nu)} \,,
\end{equation}
meaning that the traceless part of the nonmetricity tensor is completely symmetric. Using all these results, we have that \eqref{deltag12}, \eqref{contrdeltag12}, and \eqref{varconnNEW12} are identically satisfied (this means, in particular, that the equation obtained by varying w.r.t. the metric reduce to the trivial identity), and the final form for the connection reads
\begin{equation}\label{finconn12}
{\Gamma^\lambda}_{\mu \nu} = \tilde{\Gamma}^\lambda_{\phantom{\lambda} \mu \nu} + \frac{1}{12} {\delta^\lambda}_\mu \left( Q_\nu - \frac{8}{3} S_\nu \right) + \frac{1}{12} {\delta^\lambda}_\nu \left( Q_\mu + \frac{16}{3} S_\mu \right) + \frac{1}{12} \left( Q^\lambda + \frac{16}{3} S^\lambda \right) g_{\mu \nu} + \frac{1}{2} {\Omega^\lambda}_{\mu \nu} \,.
\end{equation}
Finally, exploiting the fact that the Holst action \eqref{S12} is invariant under \eqref{genclasstr} and \eqref{trconnsymm}, meaning that it is invariant under the transformation
\begin{equation}
\Gamma^{\lambda}_{\;\;\mu\nu}\mapsto \Gamma^{\lambda}_{\;\;\mu\nu}+\delta^{\lambda}_{\mu}\xi_{\nu}+\delta^{\lambda}_{\nu}\zeta_{\mu}+\zeta^{\lambda}g_{\mu\nu} + {U^\lambda}_{\mu \nu} \,,
\end{equation}
where $\xi_\nu$ and $\zeta_\mu$ are arbitrary one-form vectors and where ${U^\lambda}_{\mu \nu}$ is a completely symmetric tensor, with the choice
\begin{equation}
{U^\lambda}_{\mu \nu} = - \frac{1}{2} {\Omega^\lambda}_{\mu \nu} \,, \quad \xi_\nu = - \frac{1}{12} \left( Q_\nu - \frac{8}{3} S_\nu \right) \,, \quad \zeta_\mu = - \frac{1}{12} \left( Q_\mu + \frac{16}{3} S_\mu \right) \,,
\end{equation}
one can prove that the final form of the connection \eqref{finconn12} reduces just to the purely Levi-Civita connection, namely \eqref{finconn1LC}.

Finally, let us also mention that in the case in which one either sets the nonmetricity to zero after having varied \eqref{S12} or considers the same model \eqref{S12} with just a torsionful connection (that is, setting the nonmetricity to zero from the very beginning), the result is, again, \eqref{finconn1LC}, as expected, at this point. Thus, for the purely Holst {Metric-Affine} gravitational term as well as for the torsionful metric case the final connection reduces to the Levi-Civita one.

\subsection{Matter coupling}

We shall now wish to add some matter to {the previously analyzed} model\footnote{Here as well as in the sequel, 
we refer to `matter' as any further contribution to total action except purely gravitational terms {(according with the terminology adopted in the literature)}.} and discuss the consistency issues that may arise. Taking into account that we are in a {Metric-Affine} framework and the metric and the affine connection are independent a priori, the matter can (and will) couple to the connection as well. Having clarified this, by adding a matter part to \eqref{S1} we have
\beq\label{SEHHM}
S_{\text{EH}+\text{Holst}+\text{M}}=\frac{1}{2 \kappa} \int d^{4}x \sqrt{-g}\Big( R+\alpha \varepsilon^{\mu\nu\alpha\beta}R_{\mu\nu\alpha\beta}\Big)  +S_{\text{M}}[g,\Gamma, \Psi] \,,
\eeq
where $S_{\text{M}}$ is the matter {action} and $\Psi$ collectively denotes the matter fields.  Proceeding, we define the energy-momentum tensor in the usual way,
\beq
T_{\mu\nu}:=-\frac{2}{\sqrt{-g}}\frac{\delta S_{\text{M}}}{\delta g^{\mu\nu}} \,,
\eeq 
with $T_{\mu\nu}=T_{\nu\mu}$. Now, since matter couples also to the connection, we also define the hypermomentum tensor \cite{Hehl:1976kv}
\beq
\Delta_{\lambda}^{\;\;\;\mu\nu}:=-\frac{2}{\sqrt{-g}}\frac{\delta S_{\text{M}}}{\delta \Gamma^{\lambda}_{\;\;\;\mu\nu}} \,,
\eeq
which encodes the microscopic characteristics of matter such as spin, dilation and shear \cite{Hehl:1976kv,obukhov1993hyperfluid,obukhov1996model,Iosifidis:2020gth}. Then, varying the action \eqref{SEHHM} w.r.t. the metric and the connection, we obtain, respectively, 
\beq\label{gmnvarMatter}
R_{(\mu\nu)}-\frac{1}{2}g_{\mu\nu}R-\alpha \varepsilon_{(\nu}^{\;\;\;\alpha\beta\gamma}R_{\mu)\alpha\beta\gamma}=\kappa T_{\mu\nu} \,,
\eeq
\beq
P_{\lambda}^{\;\;\;\mu\nu}+2 \alpha \left( {-} \frac{1}{\sqrt{-g}}(2S_{\alpha}-\nabla_{\alpha})(\sqrt{-g}\varepsilon_{\lambda}^{\;\;\mu\alpha\nu})-\varepsilon_{\lambda}^{\;\;\mu\alpha\beta} S_{\alpha\beta}^{\;\;\;\;\nu} \right)= \kappa \Delta_{\lambda}^{\;\;\;\mu\nu} \,, \label{CfE} 
\eeq
where the latter is equivalent to 
\beq
P_{\lambda}^{\;\;\;\mu\nu}+2 \alpha \Big( \varepsilon^{\mu\nu\alpha\beta}Q_{\alpha\beta\lambda}+ \varepsilon^{\mu\nu\alpha\beta}S_{\alpha\beta\lambda}-\varepsilon_{\lambda}^{\;\;\nu\alpha\beta}S_{\alpha\beta}^{\;\;\;\;\mu}\Big)= \kappa \Delta_{\lambda}^{\;\;\;\mu\nu} \,. \label{CfEnew}
\eeq
Let us first concentrate on the connection field equations{,} since many conclusions can be drawn by these alone. To start with, taking the trace of \eqref{CfE} in $\mu,\lambda$ and using the fact that the gravitational part {of the action} is projective invariant, we find the constraint
\beq\label{constrHyperm}
\Delta_{\mu}^{\;\;\;\mu\nu}=0 \,,
\eeq
{namely only projective invariant matter is allowed \cite{Hehl:1976kv,1981GReGr..13.1037H}. This means that the theory is com\-pa\-ti\-ble only with matter with a vanishing dilatonic current and{, in general, one} would {then} run into  possible inconsistencies \cite{Hehl:1976kv}.\footnote{One may break the projective invariance by constraining one of the torsion or nonmetricity vectors, or a combination  of them in a certain way through Lagrange multipliers \cite{1981GReGr..13.1037H,vitagliano2011dynamics,iosifidis2019exactly} and alleviate the constraint on the trace of {the} hypermomentum.} However, both the bosonic and fermionic fields are projective invariant (see \cite{jimenez2018teleparallel} related to this point) and therefore the presence of such fields does not introduce any consistency issue.}\footnote{Whether projective invariance should be broken or not (and, thus, if it has a physical relevance) is still an open issue, whose analysis goes beyond the aim of the present paper. In fact{,} some recent investigations suggest that keeping projective invariance is key in order to avoid ghosts \cite{jimenez2019ghosts,aoki2019scalar,percacci2020new}.} One can prove, by taking traces and contractions with the Levi-Civita tensor of \eqref{CfEnew}, that also in the matter-coupled case one may exploit projective invariance and consider an appropriate gauge choice to get rid of the nonmetricity vector $\tilde{Q}_\lambda$.\footnote{Let us recall{, here,} that neither of the combinations $4\tilde{Q}^{\mu}-Q^{\mu}$, $\tilde{Q}^{\mu}-Q^{\mu}-4S^{\mu}$, and $\tilde{S}_{\mu}$ can be set to zero because all of them are projective invariant by themselves.} Moreover, it can also be shown that the presence of matter does not allow for a vanishing $Q_{\mu}$, since the latter results to be given in terms of hypermomentum traces. Let us conclude this analysis with a comment on the case in which $\alpha = \pm \frac{i}{2}$. In this particular case, from the contraction of the connection field equations \eqref{CfEnew} with the Levi-Civita tensor, one gets the following constraint on the hypermomentum currents:
\beq
	\Delta_{(2)}^{\mu}-\Delta_{(3)}^{\mu}\mp i \varepsilon^{\alpha\beta\gamma\mu}\Delta_{\alpha\beta\gamma}=0 \,,
\eeq
where we have defined $\Delta_{(2)}^{\mu}:= {\Delta^{\lambda \mu}}_{\lambda}$ and $\Delta_{(3)}^{\mu}:={\Delta^{\mu \lambda}}_{\lambda}$, along with $\Delta^{\mu}_{(1)}:={\Delta_{\lambda}}^{\lambda\mu} =0$ (which is, in fact, \eqref{constrHyperm}). We shall {refer to this type of matter as matter with a} `parity restricted hypermomentum'. It would be very interesting to find physical systems for which the latter condition {on matter fields} holds true.\footnote{For instance, it would be worth to study applications to the perfect hyperfluid discussed in \cite{Iosifidis:2020gth}.} 
Moreover, as we have previously observed, for the specific choice $\alpha=\pm \frac{i}{2}$ the gravitational part of the action results to be invariant also under connection transformations of the form \eqref{enlsymm}, where the tensor $V_{\lambda \mu \nu} $ satisfies the constraints given in \eqref{Vtensprop}. Again, this fact puts some constraints on the matter part of the theory and, in particular, it implies that the hypermomentum obeys
\beq
\Delta_{\lambda \mu \nu} = \pm {i} \varepsilon_{\lambda \mu \rho \sigma} {\Delta_\nu}^{\rho \sigma} \,, \quad \Delta_{[\lambda \mu \nu]} =0 \,.
\eeq
This last condition means that for the cases $\alpha=\pm \frac{i}{2}$ there is no room for coupling with purely fermionic matter (a detailed study of the coupling with purely fermionic matter will be given in the sequel).
Here, matter coupling certainly deserves further investigation in order to conclude whether the above restrictions on the hypermomentum are always unphysical or there are systems where these arise naturally. We conclude by mentioning that if one considers the torsionful metric case with matter (that is, if one considers vanishing nonmetricity from the very beginning) the constraint \eqref{constrHyperm} does not arise anymore. In particular, in that case one obtains
\begin{equation}
\tilde{S}^\nu = - \frac{k}{4 \alpha} \Delta^{\nu}_{(1)} \,.
\end{equation}
Nevertheless, one can also prove that other constraints on the hypermomentum arise (which, as usual, can be related to invariances of the gravitational part of the action), meaning that, again, matter results to be constrained by the gravitational part of the theory. In the sequel, we will analyze the case in which the {Metric-Affine} EH plus Holst term theory is coupled with purely fermionic matter.

\subsubsection{Purely fermionic matter}

Let us now concentrate on the case in which the {Metric-Affine} EH plus Holst term model is coupled with purely fermionic matter. In this case, the hypermomentum has only a non-vanishing totally antisymmetric part, namely
\beq
\Delta_{\alpha\mu\nu} = \Delta_{[\alpha\mu\nu]} := \hat{\Delta}_{\alpha\mu\nu} \,, \label{fer}
\eeq
with the other pieces all vanishing. Then, from the connection field equations it follows that \eqref{Q} and \eqref{Qt} hold, together with
\beq\label{StDelta}
2(1+4 \alpha^{2})\tilde{S}^{\mu}= \kappa \varepsilon^{\alpha\beta\gamma\mu} \hat{\Delta}_{\alpha\beta\gamma} \,.
\eeq
From \eqref{StDelta} we can see that for $\alpha= \pm \frac{i}{2}$ we get the constraint $\varepsilon^{\alpha\beta\gamma\mu}\hat{\Delta}_{\alpha\beta\gamma}=0$, which means, taking into account \eqref{fer},
\begin{equation}\label{hypASzero}
\hat{\Delta}_{\alpha\mu\nu} = 0 \,.
\end{equation}
Thus, as a consequence of the equations obtained from the variation w.r.t the connection with $\alpha= \pm \frac{i}{2}$ we find that that for the theory coupled with purely fermionic matter the hypermomentum vanishes on-shell. Therefore, we can conclude that the cases $\alpha=\pm \frac{i}{2}$ {lead to an inconsistency related to the vanishing of the hypermomentum when coupling with purely fermionic matter is considered}. On the other hand, for $\alpha \neq \pm \frac{i}{2}$ from \eqref{StDelta} we get
\beq
\tilde{S}^{\mu}=\frac{\kappa}{2(1+4 \alpha^{2})} \varepsilon^{\alpha\beta\gamma\mu}\hat{\Delta}_{\alpha\beta\gamma} \,,
\eeq
which shows how {matter sources torsion in this case}. Note now that in the theory currently under analysis, which is invariant under projective transformations of the connection \eqref{projectivetransf},\footnote{Indeed, following \cite{Iosifidis:2019jgi} one can prove that, due to the fact that the hypermomentum here is completely traceless (see \eqref{fer}), the matter contribution to the full action result to be invariant under projective transformations \eqref{projectivetransf} (in the matter Lagrangian, here, only the completely antisymmetric part of the contorsion contributes, giving vanishing traces).} we still have that the projective freedom can be exploited in such a way to remove a vector by choosing an appropriate gauge. Interestingly, we cannot set the combination $\tilde{Q}_{\mu}-Q_{\mu}-4 S_{\mu}$ to zero because it is projective invariant by itself.\footnote{This can be proved by considering \eqref{deltaconn-QT} together with \eqref{gennm} and \eqref{gentor}, taking contractions, and subsequently applying \eqref{projectivetransf} (see also \cite{Iosifidis:2019jgi}, where some projective invariant combinations {have been} explicitly written).} Of course the same holds true for $\tilde{S}_{\mu}$. A natural gauge would be the one for which the nonmetricity traces are removed. Considering such a gauge, we can set $\tilde{Q}_{\mu}=0$ and by means of \eqref{Qt} we also have $Q_{\mu}=0$. In addition, here the hypermomentum does not {source} neither $\Omega_{\lambda\mu\nu}$ nor $Z_{\lambda\mu\nu}$. Indeed, one can prove that we are left with \eqref{Omega} and \eqref{Zeta}. Then, we can safely set both $\Omega_{\lambda\mu\nu}$ and $Z_{\lambda\mu\nu}$ to zero.\footnote{{See} the previous discussion regarding the theory in vacuum; here, since the hypermomentum is completely antisymmetric{,} we can consistently gauge away $\Omega_{\lambda\mu\nu}$ and $Z_{\lambda\mu\nu}$ by exploiting the same transformations.} Thus, with a proper gauge-fixing, we end up with {\eqref{nonmetzero} and}
\begin{equation}
S_{\lambda} = \frac{\alpha}{2} \tilde{S}_\lambda \,,
\end{equation}
and, therefore, with
\beq
S_{\mu\nu\lambda} =\frac{\kappa}{1+4 \alpha^{2}}\left( \frac{\alpha}{6} g_{\lambda [\mu}\varepsilon_{\nu]\alpha\beta\gamma}\hat{\Delta}^{\alpha\beta\gamma} -\frac{1}{2}\hat{\Delta}_{\mu\nu\lambda} \right) \,,
\eeq
where we recall that $\hat{\Delta}_{\mu\nu\lambda}$ is completely antisymmetric.
Concluding, the final form for the connection results to be
\begin{equation}\label{connfinMfer}
{\Gamma^\lambda}_{\mu \nu} = \tilde{\Gamma}^\lambda_{\phantom{\lambda} \mu \nu} + \frac{\kappa}{2(1+4 \alpha^{2})} \left[ \frac{\alpha}{3} \left( - {\delta^\lambda}_\nu \varepsilon_{\mu \alpha \beta \gamma} + {\varepsilon^\lambda}_{\alpha \beta \gamma} g_{\mu \nu} \right) \hat{\Delta}^{\alpha \beta \gamma}  - \hat{\Delta}^{\lambda}_{\phantom{\lambda}\mu\nu} \right] \,.
\end{equation}
Let us also observe that, considering the equation obtained by varying the action w.r.t. the metric, that is \eqref{gmnvarMatter}, where now \eqref{fer} holds, and taking its $\mu,\nu$ trace, after some algebraic manipulation and using all the equations and information above, we get
\begin{equation}\label{gmnvartrMfer}
\tilde{R} = \frac{\kappa^2}{4 (1+4\alpha^2)} \hat{\Delta}_{\alpha \beta \gamma} \hat{\Delta}^{\alpha \beta \gamma} - \kappa T \,,
\end{equation}
where we have defined the trace of the energy-momentum tensor as $T:=g^{\mu \nu} T_{\mu \nu}$. {Combining \eqref{gmnvartrMfer} and \eqref{gmnvarMatter} with \eqref{fer}}, we are finally left with
\begin{equation}\label{gmvarfinMfer}
\tilde{R}_{\mu \nu} = \kappa \left( T_{\mu \nu} - \frac{1}{2} g_{\mu \nu} T \right) + \frac{\kappa}{4(1+4\alpha^2)} \hat{\Delta}_{\mu}^{\phantom{\mu} \alpha \beta} \hat{\Delta}_{\nu \alpha \beta} \,.
\end{equation}
From eqs. \eqref{gmnvartrMfer} and \eqref{gmvarfinMfer} we can see that the Ricci scalar and the Ricci tensor of the Levi-Civita connection are entirely given in terms of matter contributions, namely of the energy-momentum and hypermomentum tensors. The extra term involving the hypermomentum in \eqref{gmvarfinMfer} is a spin-spin interaction. This is easily seen as follows: The Dirac Lagrangian in the presence of torsion reads $\mathcal{L}_{D}=\mathcal{\tilde{L}}_{D}+\bar{\psi}\gamma^{[\mu}\gamma^{\nu}\gamma^{\rho]}\psi K_{\rho\mu\nu}$. Here, $\mathcal{\tilde{L}}_{D}$ is the Riemannian part and $K_{\rho\mu\nu}$ is the contorsion tensor. Now, from the definition of the hypermomentum we get $\hat{\Delta}_{\mu\nu\alpha}\propto \bar{\psi}\gamma_{[\mu\nu\alpha]}\psi$, where $\gamma_{[\mu\nu\alpha]}:=\gamma_{[\mu}\gamma_{\nu}\gamma_{\alpha]}$ and where $\gamma_\mu$ denotes the Dirac gamma matrices in four dimensions. Finally, using the relation $\gamma^{[\mu\nu\alpha]}=-i \varepsilon^{\lambda\mu\nu\alpha}\gamma_{\lambda}\gamma^{5}$, it follows that $\hat{\Delta}^{\mu\nu\alpha}\propto \varepsilon^{\lambda\mu\nu\alpha}\bar{\psi}\gamma_{\lambda}\gamma^{5}\psi$, from which $\hat{\Delta}_{\mu\nu\alpha}\hat{\Delta}^{\mu\nu\alpha}\propto (\bar{\psi}\gamma_{\mu}\gamma^{5}\psi)(\bar{\psi}\gamma^{\mu}\gamma^{5}\psi)$, as stated. Here, let us also mention that this amounts to a four-fermions interaction that originates from torsion associated with spin degrees of freedom, recently analyzed in \cite{Shaposhnikov:2020aen}. There, such an interaction has been proved to lead to a novel universal mechanism for producing singlet fermions in the Early Universe, with these fermions playing the role of dark matter particles. It would be interesting to further analyze this aspect in the models we are proposing in the current paper.

As a final comment, let us mention that, as we can see, there exists a well-defined limit $\alpha \rightarrow \infty$ for which the torsion vanishes and the theory results to be equivalent to GR. Indeed, in the limit $\alpha \rightarrow \infty$, from \eqref{connfinMfer} we find that the final form of the connection reduces to the purely Levi-Civita one, that is \eqref{finconn1LC}, and from \eqref{gmnvartrMfer} and \eqref{gmvarfinMfer} we obtain
\begin{equation}
\tilde{R}_{\mu \nu} - \frac{1}{2} g_{\mu \nu} \tilde{R} = \kappa  T_{\mu \nu} \,,
\end{equation}
with 
\begin{equation}
\tilde{R} = - \kappa T \,,
\end{equation}
which correspond to the {Einstein's} field equations. {This means that in the strong parity violating regime the non-Riemannian contributions die off and the theory reduces to GR.}

\section{{Metric-Affine} $f(R)$ gravity with Holst term}\label{fRHolst}

Let us now consider a {Metric-Affine} gravitational action consisting of the {Metric-Affine} $f(R)$ gravity contribution along with the parity violating Holst term. The total action now reads
\beq
S_{f(R)+\text{Holst}}=\frac{1}{2 \kappa} \int d^{4}x \sqrt{-g}\Big( f(R)+\alpha \varepsilon^{\mu\nu\alpha\beta}R_{\mu\nu\alpha\beta}\Big) \,, \label{S2}
\eeq
where $f(R)$ is an arbitrary function of the scalar curvature $R=g^{\mu \nu} R_{\mu \nu}\left(\Gamma\right)$, with $\Gamma$ a general affine connection.
Variation of the above w.r.t. $g^{\mu \nu}$ gives
\begin{equation}\label{varg2}
f'(R) R_{(\mu \nu)} - \frac{f(R)}{2} g_{\mu \nu} - \alpha {\varepsilon_{(\nu}}^{\alpha \beta \gamma} R_{\mu) \alpha \beta \gamma} = 0 \,,
\end{equation}
while the variation w.r.t. ${\Gamma^\lambda}_{\mu \nu}$ leads to
\begin{equation}\label{varconn2}
P_{\lambda}^{\;\;\;\mu\nu} + {\delta_\lambda}^\nu g^{\mu \sigma} \frac{\partial_\sigma f'}{f'} - g^{\mu \nu} \frac{\partial_\lambda f'}{f'} + \frac{2 \alpha}{f'} \Big( \varepsilon^{\mu\nu\alpha\beta}Q_{\alpha\beta\lambda}+ \varepsilon^{\mu\nu\alpha\beta}S_{\alpha\beta\lambda}-\varepsilon_{\lambda}^{\;\;\nu\alpha\beta}S_{\alpha\beta}^{\;\;\;\;\mu}\Big)=0 \,,
\end{equation}
where $f'=f'(R)$. 
The trace of \eqref{varg2} yields
\begin{equation}\label{tracevarg2}
f'(R) R - 2 f(R) = \alpha \varepsilon^{\mu \alpha \beta \gamma} R_{\mu \alpha \beta \gamma} \,,
\end{equation}
which can be also rewritten as 
\begin{equation}\label{tracevargfR}
\frac{f}{f'} = \frac{R}{2} - \frac{\alpha}{2f'} \varepsilon^{\mu \alpha \beta \gamma} R_{\mu \alpha \beta \gamma} \,.
\end{equation}
{Observe that substituting} \eqref{tracevarg2} into \eqref{varg2} the latter can be rewritten as
\begin{equation}
f'(R) R_{(\mu \nu)} - \frac{1}{4} \left( f'(R) R - \alpha \varepsilon^{\rho \alpha \beta \gamma} R_{\rho \alpha \beta \gamma} \right) g_{\mu \nu} - \alpha {\varepsilon_{(\nu}}^{\alpha \beta \gamma} R_{\mu) \alpha \beta \gamma} = 0 \,, 
\end{equation}
that is
\begin{equation}\label{tracelessvargfR}
R_{(\mu \nu)} - \frac{1}{4} R g_{\mu \nu} + \frac{\alpha}{4f'} \varepsilon^{\rho \alpha \beta \gamma} R_{\rho \alpha \beta \gamma} g_{\mu \nu} - \frac{\alpha}{f'} {\varepsilon_{(\nu}}^{\alpha \beta \gamma} R_{\mu) \alpha \beta \gamma} = 0 \,,
\end{equation}
which is correctly traceless in $\mu,\nu$.
Notice that if we restrict to the (trivial) {case $\alpha=0$ and also take} $f(R)=R$, from \eqref{tracevarg2} we get $R=0$ and then the metric field equations \eqref{varg2} boil down to $R_{(\mu \nu)}=0$, just as expected for the purely EH case without matter fields. Thus, one could consistently take the limit $\alpha=0$, $f(R)=R$ and arrive at the well-known result of Einstein gravity modulo projective invariance (see for instance \cite{Iosifidis:2019jgi} for details).

We can now move on to the analysis of the field equations.
Taking the $\lambda,\mu$ trace of \eqref{varconn2} we get the trivial identity, which is in accordance with the invariance of \eqref{S2} under the projective transformations of the connection given in \eqref{projectivetransf}. Considering, on the other hand, the $\lambda,\nu$ trace and the contraction with $g^{\mu \nu}$ of \eqref{varconn2}, after some algebraic manipulation we obtain
\begin{equation}\label{QfR}
Q_\lambda = \frac{8\alpha}{3 f'} \tilde{S}_\lambda - \frac{16}{3} S_\lambda + 4 \frac{\partial_{\lambda}f'}{f'} \,,
\end{equation}
\begin{equation}\label{QtfR}
\tilde{Q}_\lambda = \frac{1}{4} Q_\lambda = \frac{2\alpha}{3 f'} \tilde{S}_\lambda - \frac{4}{3} S_\lambda + \frac{\partial_{\lambda}f'}{f'} \,.
\end{equation}
Plugging \eqref{QfR} and \eqref{QtfR} into \eqref{varconn2} and contracting the resulting equation with ${\varepsilon^\lambda}_{\mu \nu \rho}$ we find
\begin{equation}\label{pdfSt}
\partial_\rho  \ln f' = - \frac{4\alpha^2+f'^2}{6\alpha f'} \tilde{S}_\rho \,.
\end{equation}
With this, we may express the above as
\begin{equation}\label{QfRnew}
Q_\lambda = - \frac{2f'}{3 \alpha} \tilde{S}_\lambda - \frac{16}{3} S_\lambda \,,
\end{equation}
\begin{equation}\label{QtfRnew}
\tilde{Q}_\lambda = \frac{1}{4} Q_\lambda = -\frac{f'}{6 \alpha} \tilde{S}_\lambda - \frac{4}{3} S_\lambda \,.
\end{equation}
Then, substituting all these results into \eqref{varconn2},  the latter becomes
\begin{equation}\label{varconnOmZeta}
\frac{2\alpha}{f'} \left[ {\varepsilon^{\mu \nu}}_{\rho \sigma} \left( {\Omega^{\rho \sigma}}_\lambda - 2 {Z_\lambda}^{\rho \sigma} \right) - 2 {\varepsilon^\nu}_{\lambda \rho \sigma} Z^{\mu \rho \sigma} \right] - {\Omega_\lambda}^{\mu \nu} - 2 {Z_\lambda}^{\mu \nu} = 0 \,.
\end{equation}
Taking {contractions} of \eqref{varconnOmZeta} with the Levi-Civita tensor{, after some algebraic manipulations we} obtain the following equations:
\begin{equation}\label{ZetafR}
Z_{\mu \nu \lambda} = \frac{12 \alpha^2 + f'^2}{4 \alpha f'} \varepsilon_{\mu \nu \rho \sigma} {Z_\lambda}^{\rho \sigma} \,,
\end{equation}
\begin{equation}\label{OmegafR}
\Omega_{\lambda \mu \nu} = - 3 \frac{4 \alpha^2 + f'^2}{f'^2} Z_{\lambda (\mu \nu)} \,.
\end{equation}
Finally, one can prove that, using \eqref{ZetafR} and \eqref{OmegafR}, eq. \eqref{varconnOmZeta} results to be identically satisfied, meaning that we do not get any further constraint.
Let us now carry on the analysis by studying separately the solutions with constant $f'$ (recall that here we are considering $\alpha$ to be a constant parameter) and the case in which $f'$ is not a constant.

Before doing that, we {remark}, here, that due to the projective invariance of the {Metric-Affine} $f(R)$ gravity action plus the Holst contribution, the coupling with matter leads again to \eqref{constrHyperm}, meaning that only projective invariant matter is allowed.

\subsection{Solutions with $f'=\text{constant}$}

We start by analyzing the cases in which $f'=\text{constant}$.
First of all, let us observe that if
\begin{equation}\label{fpC0}
f'(R) = C_0 \,,
\end{equation}
where $C_0$ is an arbitrary constant, then we get
\begin{equation}\label{fC0RplusC1}
f(R) = C_0 R + C_1 \,,
\end{equation}
where $C_1$ is an arbitrary integration constant. Moreover, we have
\begin{equation}\label{partialfpzero}
\partial_\rho f' = 0 \,.
\end{equation}

Let us mention, here, that the action \eqref{S2}, in principle, is just invariant under projective transformations \eqref{projectivetransf} (see \cite{Iosifidis:2019jgi} for details on the projective invariance of {Metric-Affine} $f(R)$ gravity). Nevertheless, if we consider \eqref{fpC0}, this automatically yields \eqref{fC0RplusC1}, and the equations obtained by considering the same from the very beginning in \eqref{S2} coincide with the equations that we obtain by considering \eqref{fpC0} and \eqref{fC0RplusC1} after having varied the action. Then, if we take \eqref{fpC0} and \eqref{fC0RplusC1} from the very beginning in \eqref{S2}, we have that the latter has the same symmetries that we have discussed in the previous section when studying the {Metric-Affine} EH + Holst theory. {Finally}, using the equations above and the symmetries of the action, after some straightforward calculations, one can prove that in all the cases in which \eqref{fpC0} holds true one is left with a final form of the connection that is purely Levi-Civita.

Concluding, we observe that the case $f'=\text{constant}$ takes us back to the results we obtained in the previous section, that is to the MAG theory with EH + Holst. Note that here, in addition, we also have a cosmological constant, so that the theory actually is EH + Holst + Cosmological constant. However, the presence of the cosmological constant does not modify in any way our previous analysis and only adds the usual extra term in the field equations for the metric. This is so because its presence does not modify in any way the connection field equations (being independent of $\Gamma$). Therefore, in the case $f'=\text{constant}$ we still get GR, but in the presence of a cosmological constant.

\subsection{Case with $f'\neq\text{constant}$}

Let us now turn our attention to the $f'\neq\text{constant}$ case. This turns out to be much more interesting then the previous one. Before proceeding, let us recall that, in this (more general) case, \eqref{S2} is invariant just under the projective transformations \eqref{projectivetransf}, meaning that there is no other invariance that one could exploit to gauge-fix non-physical degrees of freedom.

In this case, \eqref{ZetafR} yields \eqref{Zetazero} and \eqref{OmegafR} becomes \eqref{Omegazero}, that is both $Z_{\lambda \mu \nu}$ and $\Omega_{\lambda \mu \nu}$ vanish.
Let us also notice, for the sake of simplicity, that \eqref{QfR} and \eqref{QfRnew} can be recast as follows:
\begin{align}
Q_\lambda &= 4 \tilde{Q}_\lambda \,, \label{Q4Qt} \\
S_\lambda &= - \frac{3}{4} \tilde{Q}_\lambda - \frac{f'}{8\alpha} \tilde{S}_\lambda \,, \label{Stracetor}
\end{align}
where we recall that \eqref{pdfSt} holds.
Then, after some algebraic manipulation, we find that now the connection reads
\begin{equation}\label{connfinfpNOTconst}
{\Gamma^\lambda}_{\mu \nu} = \tilde{\Gamma}^\lambda_{\phantom{\lambda} \mu \nu} + \frac{1}{2} {\delta^\lambda}_\mu \tilde{Q}_\nu + \frac{1}{6} {\varepsilon^\lambda}_{\mu \nu \rho} \tilde{S}^\rho + \frac{f'}{12 \alpha} \left( g_{\mu \nu} \tilde{S}^\lambda - {\delta^\lambda}_\nu \tilde{S}_\mu \right) \,.
\end{equation}
One can then properly exploit the invariance of the action \eqref{S2} under projective transformations of the connection, \eqref{projectivetransf}, to get rid of the nonmetricity vector $\tilde{Q}_\mu$ (choosing $\xi_\nu=-\frac{1}{2}\tilde{Q}_\nu$ in \eqref{projectivetransf}). Then, from \eqref{Q4Qt} we can see that, automatically, also the nonmetricity vector $Q_\lambda$ is eliminated from the theory. In this way, we are left with just one independent vector, that is the torsion pseudo-vector $\tilde{S}_\lambda$, and the final form for the connection becomes
\begin{equation}\label{connfinfpNOTconstnew}
{\Gamma^\lambda}_{\mu \nu} = \tilde{\Gamma}^\lambda_{\phantom{\lambda} \mu \nu} + \frac{1}{6} {\varepsilon^\lambda}_{\mu \nu \rho} \tilde{S}^\rho + \frac{f'}{12 \alpha} \left( g_{\mu \nu} \tilde{S}^\lambda - {\delta^\lambda}_\nu \tilde{S}_\mu \right) \,,
\end{equation}
where $\tilde{S}_\mu$ obeys \eqref{pdfSt}, which can also be rewritten as
\begin{equation}\label{pdfStnew}
 \tilde{S}_\mu =-3 \partial_{\mu}\Big[ \arctan{\Big(\frac{f'}{2 \alpha}\Big)} \Big] \,. 
\end{equation}
Then, from \eqref{Stracetor}, taking the above discussion into account and using \eqref{pdfStnew}, for the torsion vector we get
\beq\label{pdfSnew}
S_{\mu}=\frac{3}{8}\partial_{\mu}\Big[ \ln{(f'^{2}+ 4 \alpha^{2})}\Big] \,.
\eeq
{From the last two equations  we conclude} that both the torsion vector and pseudo-vector are exact and sourced by $f'$. This fact reflects a certain similarity with the case of the so-called Palatini $f(R)$ gravity with matter {(see, for instance, \cite{Iosifidis:2019jgi} and references therein for details on this point).}\footnote{{With the term `Palatini' here we mean that the hypermomentum tensor vanishes identically, that is the matter action is independent of the connection.}}
Finally, using the information above, one can prove that eq. \eqref{tracevargfR} becomes
\begin{equation}\label{tracegfpNOTconst0}
\frac{f}{f'} = \frac{R}{2} + \frac{1}{6} \tilde{S}_\mu \tilde{S}^\mu - \frac{\alpha}{f'} \tilde{\nabla}_\mu \tilde{S}^\mu \,{.}
\end{equation}
{Given the form  for $\tilde{S}_{\mu}$ in \eqref{pdfStnew}, the latter equation}  can be seen a second order differential equation for $R$. This will become clearer in an explicit example we will discuss in the sequel. In the present case, \eqref{tracegfpNOTconst0} can be recast as follows: 
\begin{equation}\label{tracegfpNOTconst}
\frac{f}{f'} = \frac{\tilde{R}}{2} + \left( \frac{1}{12} - \frac{f'^2}{16 \alpha^2} \right) \tilde{S}_\mu \tilde{S}^{\mu} + \left(\frac{f'}{4\alpha} - \frac{\alpha}{f'} \right) \tilde{\nabla}_\mu \tilde{S}^\mu \,,
\end{equation}
where we have used the fact that now we have
\begin{equation}\label{RicciScfpNOTconst}
R = \tilde{R} - \frac{1}{2} \left( \frac{1}{3} + \frac{f'^2}{4 \alpha^2} \right) \tilde{S}_\lambda \tilde{S}^\lambda + \frac{f'}{2\alpha} \tilde{\nabla}_\lambda \tilde{S}^\lambda \,.
\end{equation}
On the other hand, \eqref{tracelessvargfR} become
\begin{equation}\label{tracelessgfpNOTconst}
R_{(\mu \nu)} - \frac{1}{4} R g_{\mu \nu} - \frac{1}{6} g_{\mu \nu} \left( \frac{1}{6} \tilde{S}_\rho \tilde{S}^\rho + \frac{\alpha}{f'} \tilde{\nabla}_\rho \tilde{S}^\rho \right) + \frac{1}{9} \tilde{S}_\mu \tilde{S}_\nu + \frac{2 \alpha}{3 f'} \tilde{\nabla}_{(\mu} \tilde{S}_{\nu)} = 0 \,,
\end{equation}
where the only independent vector appearing in \eqref{tracelessgfpNOTconst} (also in the first two terms of the latter) is just the torsion pseudo-vector $\tilde{S}_\mu$.
Thus, we can conclude that in the case $f'\neq \text{constant}$ we are left with the connection \eqref{connfinfpNOTconstnew} where the torsion pseudo-vector $\tilde{S}_\mu$ is exact and obeys \eqref{pdfStnew}, together with the above equations \eqref{tracegfpNOTconst} and \eqref{tracelessgfpNOTconst}.
Let us now observe that one can integrate \eqref{pdfStnew}, obtaining
\begin{equation}\label{fpStint}
\arctan \left( \frac{f'}{2\alpha} \right) = \Phi \,,
\end{equation}
where we have defined
\begin{equation}\label{Phidefinition}
\Phi :=  - \frac{1}{3} \int \tilde{S}_\mu dx^\mu + \text{constant} \,.
\end{equation}
From \eqref{fpStint} it follows that
\begin{equation}\label{fpform}
f' = 2 \alpha \tan \Phi \,.
\end{equation}
Using $\frac{df}{dR}=\frac{df}{d\Phi}\frac{d\Phi}{dR}$, we can see that \eqref{fpform} implies 
\begin{equation}\label{fform}
f = - 2 \alpha \ln \left( \cos \Phi \right) \,.
\end{equation}
{Using} \eqref{fform} into \eqref{pdfSnew}, the latter yields
\begin{equation}\label{SPhi0}
\tilde{S}_\mu = - 3 \partial_\mu \left[\arctan\left(\tan \Phi \right) \right] \,.
\end{equation}
Assuming, consistently, that $R$ takes values in $\mathbb{R}$, then we find, from \eqref{fpform}, that if we take
\begin{equation}\label{rangePhi}
\Phi \in \left( - \frac{\pi}{2} , \frac{\pi}{2} \right) \,,
\end{equation}
this range for $\Phi$ automatically maps to the whole real line. Thus, the choice \eqref{rangePhi} does not introduce any {additional} constraint in the present theory. With the choice \eqref{rangePhi}, eq. \eqref{SPhi0} becomes
\begin{equation}\label{SPhi}
\tilde{S}_\mu = - 3 \partial_\mu \Phi \,.
\end{equation}
Here, notice also that the same would have been obtained by simply {substituting} the left-hand side of \eqref{fpStint} {written} in terms of $\Phi$ into \eqref{pdfSnew}{,} {confirming that what we are performing is consistent and well-defined by construction.}
Consequently, eq. \eqref{tracegfpNOTconst0} yields,
\begin{equation}\label{RSeqwewillalsouse}
- \left( \cot \Phi \right) \ln \left( \cos \Phi \right) = \frac{R}{2} + \frac{3}{2 \sqrt{-g}} \left( \cot \Phi \right) \partial_\mu \left( \sqrt{-g} \partial^\mu \Phi \right) + \frac{3}{2} \partial_\mu \Phi \partial^\mu \Phi \,.
\end{equation}
Using the fact that \eqref{RicciScfpNOTconst}, 
{which is the decomposition of $R$ in terms of $\tilde{R}$ and non-Riemannian contributions,}
now reads as follows:
\begin{equation}\label{nonRiemexpofR}
\begin{split}
R & = \tilde{R} - \frac{3}{\sqrt{-g}} \left(\tan \Phi \right) \partial_\mu \left( \sqrt{-g} \partial^\mu \Phi \right) + \frac{3}{2} \left(2-3 \left( \sec \Phi \right)^2 \right) \partial_\mu \Phi \partial^\mu \Phi \\
& = \tilde{R} - \frac{3}{\sqrt{-g}} \left(\tan \Phi \right) \partial_\mu \left( \sqrt{-g} \partial^\mu \Phi \right) - \frac{3}{2} \left(1+3 \left( \tan \Phi \right)^2 \right) \partial_\mu \Phi \partial^\mu \Phi \,,
\end{split}
\end{equation}
we find that \eqref{RSeqwewillalsouse} leads to
\begin{equation}\label{tildeRFpNOTconst}
\tilde{R} = - 2 \left( \cot \Phi \right) \ln \left( \cos \Phi \right) + \frac{3}{\sqrt{-g}} \left( \tan \Phi - \cot \Phi \right) \partial_\mu \left( \sqrt{-g} \partial^\mu \Phi \right) - \frac{3}{2} \left(1 - 3 \left( \tan \Phi \right)^2 \right) \partial_\mu \Phi \partial^\mu \Phi \,,
\end{equation}
which gives $\tilde{R}$ in terms of functions of $\Phi$ and its derivatives, and one can also prove that eq. \eqref{tracelessgfpNOTconst} becomes
\begin{equation}\label{geneqPhitildeRmunu}
\begin{split}
& \tilde{R}_{\mu \nu} - \frac{1}{4} g_{\mu \nu} \tilde{R} + \frac{1}{8} g_{\mu \nu} \left[ \left(\sec \Phi \right)^2 \partial_\rho \Phi \partial^\rho \Phi + \frac{2}{\sqrt{-g}} \left( \csc \Phi \right) \left( \sec \Phi \right) \partial_\rho \left( \sqrt{-g} \partial^\rho \Phi \right) \right] \\
& - \frac{1}{2} \left(\sec \Phi \right)^2 \partial_\mu \Phi \partial_\nu \Phi - \left( \csc \Phi \right) \left( \sec \Phi \right) \tilde{\nabla}_\mu \partial_\nu \Phi = 0 \,.
\end{split}
\end{equation}
One can also plug the expression for $\tilde{R}$ given in \eqref{tildeRFpNOTconst} into \eqref{geneqPhitildeRmunu}, ending up with $\tilde{R}_{\mu \nu}$ in terms  of $\Phi$ and its derivatives, namely
\begin{equation}\label{RmunuPhitocompare}
\begin{split}
\tilde{R}_{\mu \nu} & = g_{\mu \nu} \Bigg [ - \frac{1}{2} \left( \cot \Phi \right) \ln \left(\cos \Phi \right) - \left( \frac{1}{2} - \left( \tan \Phi \right)^2 \right) \partial_\rho \Phi \partial^\rho \Phi \\
& + \frac{1}{\sqrt{-g}} \left( \frac{1}{2} \tan \Phi - \cot \Phi \right) \partial_\rho \left( \sqrt{-g} \partial^\rho \Phi \right) \Bigg ] + \frac{1}{2} \left(\sec \Phi \right)^2 \partial_\mu \Phi \partial_\nu \Phi \\
& + \left( \csc \Phi \right) \left( \sec \Phi \right) \tilde{\nabla}_\mu \partial_\nu \Phi \,.
\end{split}
\end{equation}
We will see later some explicit examples with different assumptions on the form of $f(R)$. {Notice that  the right-hand side of the above suggests that there is  some additional scalar mode at play here. This is indeed the case{,} as we formally prove in what follows.}
Let us conclude this analysis by {saying} that using all the relations we have obtained so far we also have
\begin{equation}\label{HolstPhi}
\alpha \varepsilon^{\mu \nu \rho \sigma} R_{\mu \nu \rho \sigma} = - 6 \alpha \left[ \left(\tan \Phi \right) \partial_\mu \Phi \partial^\mu \Phi + \frac{1}{\sqrt{-g}} \partial_\mu \left( \sqrt{-g} \partial^\mu \Phi \right) \right] \,.
\end{equation}
Eq. \eqref{HolstPhi}{, together with the other results collected above,} will be particularly useful in the sequel to prove an intriguing equivalence that we are going to provide.

\subsubsection{On the degrees of freedom and equivalence with a metric torsionless Scalar-Tensor theory}

As it may be already evident by now, in the case in which $f'\neq \text{constant}$ the theory seems to carry one more scalar degree of freedom with respect to GR. This is no other than the scalar mode $\Phi$ and its very appearance in \eqref{geneqPhitildeRmunu} indicates that this {should indeed be} the case. {In the sequel, we formally prove this.} Notice that the equivalence between {Metric-Affine} $f(R)$ gravity plus Holst term (in the general case $f' \neq \text{constant}$) and metric torsionless Scalar-Tensor theory we are going to discuss is an on-shell equivalence (meaning that the two theories exhibit the same dynamics) which holds true after having gauged away the (non-physical) nonmetricity degrees of freedom {from} the {Metric-Affine} $f(R)$ gravity theory plus Holst term.
 
Following the standard $f(R)$ equivalence {prescription \cite{sotiriou2006f,Sotiriou:2008rp}}, we introduce an auxiliary {scalar} field $\chi$ and consider the action
\beq
 S=\frac{1}{2\kappa}\int d^{4}x  \sqrt{-g}\Big[ f(\chi)+f'(\chi)(R-\chi)+\alpha \varepsilon^{\alpha\beta\mu\nu}R_{\alpha\beta\mu\nu} \Big] \,, \label{equiv}
 \eeq
where $f'(\chi)$ represents the derivative of $f(\chi)$ w.r.t. $\chi$. Variation of \eqref{equiv} w.r.t. $\chi$ leads to 
\begin{equation}\label{fsecRchi}
f''(\chi) \left(R-\chi\right) = 0 \,.
\end{equation}
Therefore, we have that if $f''(\chi)\neq 0$ then $\chi=R$, which, when {substituted} back into \eqref{equiv}, establishes the equivalence of the latter with \eqref{S2}. Now, in the usual manner (see \cite{sotiriou2006f,Sotiriou:2008rp}) we set $f'(\chi)=\Psi$ and suppose that this can be inverted to give $\chi=\chi(\Psi)$. With this identification the action \eqref{equiv} takes the form
\beq\label{equivnew}
 S=\frac{1}{2\kappa}\int d^{4}x  \sqrt{-g}\Big[ \Psi R+V(\Psi)+\alpha \varepsilon^{\mu \nu \rho \sigma} R_{\mu \nu \rho \sigma} \Big] \,,
\eeq
where, as usual \cite{Sotiriou:2008rp}, we have defined the potential $V(\Psi):=f(\chi(\Psi))-\Psi \chi(\Psi)$. Now, notice that on-shell we have $f'(\chi)=f'(R)$, and, therefore, using \eqref{fpform}, {we get} 
\begin{equation}\label{PsiPhi}
\Psi=2\alpha \tan{\Phi} \,,
\end{equation}
which (under the consistent assumptions previously discussed) implies $\Phi = \arctan\left(\frac{\Psi}{2\alpha} \right)$. Then, using the latter into \eqref{HolstPhi}, we obtain
\begin{equation}\label{HolstPhinew}
\alpha \varepsilon^{\mu \nu \rho \sigma} R_{\mu \nu \rho \sigma} = -\frac{3}{4\alpha^{2}}\frac{\Psi}{\left(1+\left(\frac{\Psi}{2\alpha} \right)^{2} \right)^{2}}\partial_{\mu}\Psi \partial^{\mu}\Psi - 3 \tilde{\nabla}_\mu \left[ \frac{\partial^\mu \Psi}{1+\left(\frac{\Psi}{2\alpha} \right)^2} \right] \,. 
\end{equation}
Finally, plugging \eqref{HolstPhinew} back into the action \eqref{equivnew}, disregarding the surface term, we arrive at
\beq\label{equivnew1}
 S=\frac{1}{2\kappa}\int d^{4}x  \sqrt{-g}\left[ \Psi R+V(\Psi)-\frac{3}{4\alpha^{2}}\frac{\Psi}{\left(1+\left(\frac{\Psi}{2\alpha} \right)^{2} \right)^{2}}\partial_{\mu}\Psi \partial^{\mu}\Psi \right] \,.
\eeq
At this point, this action looks like the {one} for a Scalar-Tensor theory. Nevertheless, let us recall that here $R$ is the Ricci scalar of the affine connection ${\Gamma^\lambda}_{\mu \nu}$, meaning that it also contains non-Riemannian contributions that will be now functions of $\Psi$. Using the decomposition of $R$ given in \eqref{nonRiemexpofR} one can prove that we are left with
\beq\label{sqrtgPsiR}
\sqrt{-g}\Psi R= \sqrt{-g}\Psi \tilde{R}+\sqrt{-g} \left[ 3 \left( - \frac{\Psi^2}{4 \alpha^2} \omega^{2} +\frac{3 \omega}{4 \alpha}\right) \right] \Psi \partial_{\mu}\Psi \partial^{\mu}\Psi -\frac{3}{2 \alpha}\partial_{\mu}\left[\sqrt{-g}\Psi^{2} \left(\partial^{\mu}\Phi \right) \omega \right] \,,
\eeq 
where we have defined 
\beq
\omega :=\frac{1}{2\alpha}\frac{1}{\left( 1+\frac{\Psi^{2}}{4\alpha^{2}} \right)} \,.
\eeq
{Substituting} \eqref{sqrtgPsiR} into the action \eqref{equivnew1}, discarding the surface term (that is the last term in the right-hand side of \eqref{sqrtgPsiR}), we finally get the equivalent theory
\beq\label{equivfin}
 S=\frac{1}{2\kappa}\int d^{4}x  \sqrt{-g} \left[ \Psi \tilde{R}+V(\Psi)+\frac{3}{2}\frac{\Psi}{(4\alpha^{2}+\Psi^{2})}\partial_{\mu}\Psi\partial^{\mu}\Psi \right] \,,
\eeq
which is, indeed, a metric torsionless Scalar-Tensor theory,\footnote{{See \cite{teyssandier1983cauchy,sotiriou2006f,Iosifidis:2019jgi}, where the equivalence between other $f(R)$ gravity theories and Scalar-Tensor models has been discussed.}} that is the theory propagates an additional scalar degree of freedom compared to GR. We have therefore arrived to the {remarkable} result that the vacuum {Metric-Affine} $f(R)$ gravity plus Holst term theory is equivalent to a metric torsionless Scalar-Tensor theory.

{Let us now analyze the field equations of theory \eqref{equivfin}. Taking the variation of the latter w.r.t. $\Psi$, we obtain}
\begin{equation}\label{varPsi}
\tilde{R} + V'(\Psi) + \frac{3}{4\alpha^2+\Psi^2} \left( - \Psi \tilde{\nabla}_\mu \partial^\mu \Psi + \frac{1}{4\alpha^2+\Psi^2} \Psi^2 \partial_\mu \Psi \partial^\mu \Psi - \frac{1}{2} \partial_\mu \Psi \partial^\mu \Psi \right) = 0 \,.
\end{equation}
{On the other hand, by varying the same action w.r.t. $g^{\mu \nu}$ we get}
\begin{equation}\label{vargmunusct}
\begin{split}
& \tilde{R}_{\mu \nu}  + g_{\mu \nu} \left( - \frac{1}{2} \tilde{R} - \frac{1}{2} \frac{V(\Psi)}{\Psi} + \frac{\tilde{\nabla}_\rho \partial^\rho \Psi}{\Psi} - \frac{3}{4\left(4\alpha^2+\Psi^2\right)} \partial_\rho \Psi \partial^\rho \Psi \right) \\
& + \frac{3}{2\left(4\alpha^2+\Psi^2\right)} \partial_\mu \Psi \partial_\nu \Psi - \frac{\tilde{\nabla}_\nu \left(\partial_\mu \Psi \right)}{\Psi} = 0 \,,
\end{split}
\end{equation}
whose trace yields
\begin{equation}\label{tracevargmunusct}
\tilde{R} + \frac{2}{\Psi} V(\Psi) - \frac{3}{\Psi} \tilde{\nabla}_\mu \partial^\mu \Psi + \frac{3}{2\left(4\alpha^2+\Psi^2\right)} \partial_\mu \Psi \partial^\mu \Psi = 0 \,.
\end{equation}
Plugging \eqref{tracevargmunusct} back into \eqref{vargmunusct} we find
\begin{equation}\label{tracelessvargmunusct}
\tilde{R}_{\mu \nu} + \frac{1}{2} g_{\mu \nu} \left( \frac{V(\Psi)}{\Psi} - \frac{\tilde{\nabla}_\rho \partial^\rho \Psi}{\Psi} \right) + \frac{3}{2\left(4\alpha^2+\Psi^2\right)} \partial_\mu \Psi \partial_\nu \Psi - \frac{\tilde{\nabla}_\nu \left(\partial_\mu \Psi \right)}{\Psi} = 0  \,,
\end{equation}
while from comparison of \eqref{tracevargmunusct} with \eqref{varPsi} we are left with
\begin{equation}\label{VVprime}
\frac{-2 V(\Psi)}{\Psi} + 3 \left[ \left(\frac{1}{\Psi} - \frac{\Psi}{4\alpha^2 + \Psi^2} \right) \tilde{\nabla}_\mu \partial^\mu \Psi + \left( \frac{\Psi^2}{\left(4\alpha^2+\Psi^2\right)^2} - \frac{1}{4\alpha^2+\Psi^2} \right) \partial_\mu \Psi \partial^\mu \Psi \right] + V'(\Psi) = 0 \,,
\end{equation} 
where this last equation determines the dynamics of $\Psi$.
Observe, here, that the field equations for $\Psi$ are second order ones. Finally, rewriting all the information above in terms of $\Phi$,\footnote{Using, in particular, {eqs.} \eqref{fpform} and \eqref{nonRiemexpofR}, together with \eqref{PsiPhi} and the fact that we are led to $V'(\Psi(\Phi))=-\chi=-R$.} one can prove, after some algebraic manipulation, that the equations of motion of the metric torsionless Scalar-Tensor theory \eqref{equivfin} coincide with those of the vacuum {Metric-Affine} $f(R)$ gravity plus Holst term theory previously analyzed, confirming the on-shell equivalence between the two theories. Notice, in particular, that we find out that eq. \eqref{VVprime} reduces to the trivial identity and that the description of the dynamics of the scalar degree of freedom is transferred to the equations of the {Metric-Affine} $f(R)$ gravity plus Holst term model.

It is also worth stressing out the in the small $\alpha$ limit the Scalar-Tensor theory actually becomes a Brans-Dicke theory with Brans-Dicke parameter $\omega_{0}=-\frac{3}{2}$, as it can be easily seen by neglecting the $\alpha^{2}$ term appearing in \eqref{equivnew1}.\footnote{{See} \cite{sotiriou2006f,Sotiriou:2008rp} for the equivalence between generic Palatini $f(R)$ theories of gravity with matter and Brans-Dicke gravity. {Note} that the aforementioned Palatini $f(R)$ gravity theories with matter have been proved \cite{sotiriou2006f,Sotiriou:2008rp} to be equivalent to Brans-Dicke gravity with {Brans-Dicke} parameter the same parameter $\omega_{0}=-\frac{3}{2}$, and the {this holds true in the case in which either torsion or nonmetricity vanishes as well}.}

Let us finally mention that one could also study how matter with vanishing hypermomentum affects this result. We aleady know that {Metric-Affine} $f(R)$ gravity plus matter with no hypermomentum is equivalent to a specific Brans-Dicke theory (see \cite{sotiriou2006f,sotiriou2007metric}). {In the case under analysis, we expect things to be considerably more involved and leave this study for a future work.}

\subsection{Comments on the torsionful metric case}

Here we make some comments of the torsionful metric case. First of all, let us observe {that} if we set the nonmetricity tensor $Q_{\lambda \mu \nu}$ to zero in the theory above after variation of the action \eqref{S2}, we are led to the same results obtained in the torsionful nonmetric case.

On the other hand, one could consider vanishing nonmetricity from the very beginning in \eqref{S2}. The introduction of the Holst term in an $f(R)$ theory of gravity with torsion has been previously considered in \cite{Bombacigno:2018tih}, but there the BI parameter has been promoted to a field.\footnote{{In \cite{Bombacigno:2018tih}}, the authors called $\beta(x^\mu)$ the the reciprocal of the Immirzi field that couples to the Riemann tensor by means of the completely antisymmetric tensor, that is the Levi-Civita tensor.} We will study the case where the BI parameter is promoted to a field in a future work [some work is now in progress on this point], while for the moment we shall focus on $\alpha$ being a constant parameter.

Thus, in the case in which one considers \eqref{nonmetzero} from the very beginning in \eqref{S2}, the variation of the action w.r.t. the connection yields
\begin{equation}\label{varconnfRonlyT}
\frac{4}{3} g^{\mu \nu} S_\lambda - \frac{1}{3} {\varepsilon_\lambda}^{\mu \nu \rho} \tilde{S}_\rho - 2 {Z_\lambda}^{\mu \nu} + {\delta_\lambda}^\nu g^{\mu \sigma} \frac{\partial_\sigma f'}{f'} - g^{\mu \nu} \frac{\partial_\lambda f'}{f'} + \frac{4 \alpha}{f'} \varepsilon^{\mu \nu \alpha \beta} S_{\alpha \beta \lambda} =0 \,,
\end{equation}
where we have also written explicitly the Palatini tensor with vanishing nonmetricity exploiting \eqref{Palatiniexpl}.
Taking the $\lambda,\mu$ trace of \eqref{varconnfRonlyT} we get \eqref{Stzero} (we discard, as usual, the trivial case $\alpha=0$). {Substituting} \eqref{Stzero} into \eqref{varconnfRonlyT} and taking the $\lambda,\nu$ trace of the resulting equation we get
\begin{equation}\label{Spdf}
S_\mu = \frac{3}{4} \frac{\partial_\mu f'}{f'} \,,
\end{equation}
while, using also \eqref{Spdf} and taking the contraction with the Levi-Civita tensor of the aforementioned resulting equation, we find
\begin{equation}\label{pdfzero}
\frac{\partial_\mu f'}{f'^2}= 0 \,.
\end{equation}
Observe that this is a rather strong constraint, meaning that the only solutions admitted here are the ones with $f'=\text{constant}$. We will come back to this point at the end of this analysis.
Now, plugging \eqref{pdfzero} into \eqref{Spdf}, we are led to
\begin{equation}
S_\mu = 0 \,.
\end{equation}
After some algebraic manipulation on the remaining equation, we find \eqref{Zetazero}. Thus, we are left with
\begin{equation}
{S_{\mu \nu}}^\lambda = 0 \,,
\end{equation}
meaning that the torsion vanishes and that the final form of the connection reduces to the Levi-Civita one.
Furthermore, one can prove that the equations obtained by varying the action w.r.t. the metric yield
\begin{equation}\label{vargfRonlyT}
\frac{f}{f'} = \frac{\tilde{R}}{2} \,, \quad \tilde{R}_{\mu \nu} - \frac{1}{4} g_{\mu \nu} \tilde{R} = 0 \,.
\end{equation}
Finally, as we have already mentioned, looking at \eqref{pdfzero}, we can deduce that $f'=\text{constant}$, that is \eqref{fpC0}, which implies \eqref{fC0RplusC1}. Therefore, we conclude that there is no such thing as metric torsionful $f(R)$ + Holst gravity, since $f(R)$ is forced to be linear in $R$. Namely, the theory is consistent only for the choice $f(R)=C_{0}R+C_{1}$. In this latter case, the theory turns out {to be equivalent} to vacuum GR {in the presence of a cosmological constant}.

\subsection{An explicit example: $f(R) = R + \varepsilon R^2$ plus Holst term}

Let us now consider, as an explicit example, the case in which one has {the {Metric-Affine} $f(R)$ gravity plus Holst term theory} with
\begin{equation}\label{fex1}
f(R)=R+ \varepsilon R^2 \,,
\end{equation}
{where $\varepsilon$ is a constant parameter with dimensions of inverse mass squared (or{,} equivalently, squared length).}
Here we immediately get
\begin{equation}\label{fpex1}
f'(R) = 1 + 2 \varepsilon R \,.
\end{equation}
In the following analysis, we discard the case $R=\text{constant}$, since it would just {automatically} lead to a purely Levi-Civita connection.
Thus, let us carry on our study by considering $R\neq\text{constant}$, from which we have $f'\neq\text{constant}$ and $f\neq\text{constant}$. We have that eq. \eqref{pdfStnew} now becomes
\begin{equation}\label{Stfpex1}
\tilde{S}_\mu = - 3 \partial_\mu \left[ \arctan \left( \frac{1+2\varepsilon R}{2 \alpha} \right) \right] \,,
\end{equation}
where we recall that $\tilde{S}_\rho$ is the only independent (pseudo-)vector that remains in our theory (see the discussion on projective invariance previously done), while from \eqref{pdfSnew} we get
\begin{equation}\label{Sfpex1}
S_\mu = \frac{3}{8} \partial_\mu \left[ \ln \left( \left(1+2 \varepsilon R \right)^2 + 4 \alpha^2 \right) \right] \,.
\end{equation} 
{Moreover, we find} that \eqref{tracegfpNOTconst0} becomes
\begin{equation}\label{RStotex1}
R = \frac{1+2 \varepsilon R}{3} \tilde{S}_\mu \tilde{S}^\mu - 2 \alpha \tilde{\nabla}_\mu \tilde{S}^\mu \,,
\end{equation}
while \eqref{tracelessgfpNOTconst} yields
\begin{equation}\label{Rmunutotex1}
R_{(\mu \nu)} - \frac{1}{4} R g_{\mu \nu} - \frac{1}{6} g_{\mu \nu} \left[ \frac{1}{6} \tilde{S}_\rho \tilde{S}^\rho + \frac{\alpha}{1 + 2 \varepsilon R} \tilde{\nabla}_\rho \tilde{S}^\rho \right] + \frac{1}{9} \tilde{S}_\mu \tilde{S}_\nu + \frac{2 \alpha}{3 (1 + 2 \varepsilon R)} \tilde{\nabla}_{(\mu} \tilde{S}_{\nu)} = 0 \,.
\end{equation}
Let us also mention that from \eqref{RicciScfpNOTconst} now we get
\begin{equation}\label{RSx1}
R = \tilde{R} - \frac{1}{2} \left( \frac{1}{3} + \frac{(1+2 \varepsilon R)^2}{4 \alpha^2}  \right) \tilde{S}_\lambda \tilde{S}^\lambda + \frac{1+2 \varepsilon R}{2\alpha} \tilde{\nabla}_\lambda \tilde{S}^\lambda \,,
\end{equation}
where $\tilde{S}_\mu$ is exact and given by \eqref{Stfpex1}.

Observe that integrating \eqref{Stfpex1} we get 
\begin{equation}
\arctan \left(\frac{1+2 \varepsilon R}{2 \alpha} \right) = \Phi \,,
\end{equation}
where $\Phi$ is defined in \eqref{Phidefinition}.
Then, we are left with 
\begin{equation}\label{Rinex1}
R = \frac{1}{\varepsilon} \left( \alpha \tan \Phi - \frac{1}{2} \right) \,.
\end{equation}
Thus, we find that \eqref{Stfpex1} yields \eqref{SPhi}.
{Substituting} {the above} into \eqref{RStotex1} and \eqref{Rmunutotex1} we obtain, respectively,
\begin{equation}\label{Reqfinex1}
\frac{1}{\varepsilon} \left(- \frac{1}{2} + \alpha \left(\tan \Phi \right) \right) = \frac{6 \alpha}{\sqrt{-g}} \partial_\mu \left( \sqrt{-g} \partial^\mu \Phi \right) + 6 \alpha \left( \tan \Phi \right) \partial_\mu \Phi \partial^\mu \Phi \,,
\end{equation}
which is an equation for $\Phi$, and \eqref{geneqPhitildeRmunu}.
Let us finally mention that {we can also write}
\begin{equation}\label{RSx1fin}
\tilde{R} = {\frac{1}{\varepsilon} \left( \alpha \tan \Phi - \frac{\cot \Phi}{4\alpha} \right)} + \frac{3}{\sqrt{-g}} \left( \tan \Phi - \cot \Phi \right) \partial_\mu \left( \sqrt{-g} \partial^\mu \Phi \right) - \frac{3}{2} \left(1 - 3 \left( \tan \Phi \right)^2 \right) \partial_\mu \Phi \partial^\mu \Phi \,,
\end{equation}
where the Ricci scalar $\tilde{R}$ of the Levi-Civita connection is given in terms of functions of $\Phi$ and its derivatives. One can also plug \eqref{RSx1fin} into \eqref{geneqPhitildeRmunu} to get $\tilde{R}_{\mu \nu}$ completely in terms of functions of $\Phi$ and its derivatives, that is
\begin{equation}
\begin{split}
\tilde{R}_{\mu \nu} & = g_{\mu \nu} \Bigg [ {\frac{1}{4\varepsilon} \left( \alpha \tan \Phi - \frac{\cot \Phi}{4\alpha} \right)} - \left( \frac{1}{2} - \left( \tan \Phi \right)^2 \right) \partial_\rho \Phi \partial^\rho \Phi \\
& + \frac{1}{\sqrt{-g}} \left( \frac{1}{2} \tan \Phi - \cot \Phi \right) \partial_\rho \left( \sqrt{-g} \partial^\rho \Phi \right) \Bigg ] + \frac{1}{2} \left(\sec \Phi \right)^2 \partial_\mu \Phi \partial_\nu \Phi \\
& + \left( \csc \Phi \right) \left( \sec \Phi \right) \tilde{\nabla}_\mu \partial_\nu \Phi = 0 \,.
\end{split}
\end{equation}
Furthermore, following the previous discussion on the remaining scalar degree of freedom, one can prove that the theory is on-shell equivalent to a metric torsionless Scalar-Tensor model with potential $V(\Psi)=\frac{1}{4 \varepsilon} \left(1 + 2 \Psi -\Psi^2 \right)$, where $\Psi$ is given by \eqref{PsiPhi}. 
Having studied this explicit example, we can now proceed by analyzing the restriction to the case in which $f(R)=R^2${, where we will also give an application to homogeneous cosmology}.

\subsection{Restriction to $f(R)=R^2$ plus Holst term}

Here we restrict ourselves to the case in which
\begin{equation}\label{fex2}
f(R)=R^2 \,,
\end{equation}
yielding
\begin{equation}\label{fpex2}
f'(R) = 2R \,.
\end{equation}
Let us mention that with the choice \eqref{fex2}, the $f(R)$ term in the action \eqref{S2} results to be invariant under conformal transformations of the metric tensor (as defined in \cite{Iosifidis:2019jgi}),
\begin{equation}\label{conftr}
g_{\mu \nu} \mapsto e^{2 \Omega} g_{\mu \nu} \,, \quad {\Gamma^\lambda}_{\mu \nu} \mapsto {\Gamma^\lambda}_{\mu \nu} \,,
\end{equation}
where $\Omega$ is a scalar function. Indeed, under \eqref{conftr} we have
\begin{equation}\label{tansfobjunderconf}
d^4 x \sqrt{-g} \mapsto d^4 x e^{4\Omega} \sqrt{-g} \,, \quad {R^\lambda}_{\mu \nu \rho} \mapsto {R^\lambda}_{\mu \nu \rho}  \,, \quad R_{\mu \nu} \mapsto R_{\mu \nu} \,, \quad R \mapsto R e^{-2\Omega} \,, 
\end{equation}
and one can clearly see that the $f(R)$ term in \eqref{S2} with the choice \eqref{fex2} is invariant. Nevertheless, the complete action \eqref{S2} is not invariant under \eqref{conftr}, due to the fact that the Holst term is not.
We now proceed with our analysis. Again, we discard the case $R=\text{constant}$. Now, eqs. \eqref{pdfStnew} and \eqref{pdfSnew} respectively boil down to
\begin{equation}\label{Stfpe21}
\tilde{S}_\mu = - 3 \partial_\mu \left[ \arctan \left( \frac{R}{\alpha} \right) \right] \,,
\end{equation}
\begin{equation}\label{Sfpe21}
S_\mu = \frac{3}{8} \partial_\mu \left[ \ln \left( 4 \left( R^2 + \alpha^2 \right) \right) \right] \,.
\end{equation}
On the other hand, \eqref{tracegfpNOTconst0} becomes
\begin{equation}\label{tracegex2}
\tilde{\nabla}_\mu \tilde{S}^\mu - \frac{R}{3 \alpha} \tilde{S}_\mu \tilde{S}^\mu = 0 \,{.}
\end{equation}
Moreover, eq. \eqref{tracelessgfpNOTconst} yields
\begin{equation}
R_{(\mu \nu)} - \frac{1}{4} R g_{\mu \nu} - \frac{1}{6} g_{\mu \nu} \left( \frac{1}{6} \tilde{S}_\rho \tilde{S}^\rho + \frac{\alpha}{2R} \tilde{\nabla}_\rho \tilde{S}^\rho \right) + \frac{1}{9} \tilde{S}_\mu \tilde{S}_\nu + \frac{\alpha}{3R} \tilde{\nabla}_{(\mu} \tilde{S}_{\nu)} = 0 \,,
\end{equation}
which, exploiting \eqref{tracegex2}, can {be also} rewritten as
\begin{equation}\label{Rmunupenultimateex2}
R_{(\mu \nu)} - \frac{1}{4} R g_{\mu \nu} - \frac{1}{18} g_{\mu \nu} \tilde{S}_\rho \tilde{S}^\rho + \frac{1}{9} \tilde{S}_\mu \tilde{S}_\nu + \frac{\alpha}{3R} \tilde{\nabla}_{(\mu} \tilde{S}_{\nu)} = 0 \,.
\end{equation}
Finally, \eqref{RicciScfpNOTconst} now reads
\begin{equation}\label{RSex2}
R = \tilde{R} - \frac{1}{2} \left( \frac{1}{3} + \frac{R^2}{\alpha^2} \right) \tilde{S}_\lambda \tilde{S}^\lambda + \frac{R}{\alpha} \tilde{\nabla}_\lambda \tilde{S}^\lambda \,,
\end{equation}
where, as we can see from \eqref{Stfpe21}, the torsion pseudo-vector $\tilde{S}_\mu$ is {exact}.

Let us now observe, on the same lines of what we have previously done, that we can integrate \eqref{Stfpe21} to obtain
\beq\label{goodform}
R=\alpha \tan{\Phi} \,,
\eeq
where $\Phi$ is defined as in \eqref{Phidefinition}, together with \eqref{SPhi}.
{Thus, one can prove that} eq. \eqref{tracegex2} becomes
\beq
\partial_{\mu}(\sqrt{-g}\partial^{\mu}\Phi)+\sqrt{-g} \left(\tan{\Phi} \right)\partial^{\mu}\Phi \partial_{\mu}\Phi=0 \,. \label{dif}
\eeq
Then, we get once again \eqref{geneqPhitildeRmunu}, while eq. \eqref{RSex2} becomes
\begin{equation}\label{tileRex2fin}
\tilde{R} = \alpha \tan \Phi + \frac{3}{\sqrt{-g}} \left( \tan \Phi - \cot \Phi \right) \partial_\mu \left( \sqrt{-g} \partial^\mu \Phi \right) - \frac{3}{2} \left(1 - 3 \left( \tan \Phi \right)^2 \right) \partial_\mu \Phi \partial^\mu \Phi \,.
\end{equation}
Moreover, {we get}
\begin{equation}
\begin{split}
\tilde{R}_{\mu \nu} & = g_{\mu \nu} \Bigg [ \frac{\alpha}{4} \tan \Phi - \left( \frac{1}{2} - \left( \tan \Phi \right)^2 \right) \partial_\rho \Phi \partial^\rho \Phi \\
& + \frac{1}{\sqrt{-g}} \left( \frac{1}{2} \tan \Phi - \cot \Phi \right) \partial_\rho \left( \sqrt{-g} \partial^\rho \Phi \right) \Bigg ] + \frac{1}{2} \left(\sec \Phi \right)^2 \partial_\mu \Phi \partial_\nu \Phi \\
& + \left( \csc \Phi \right) \left( \sec \Phi \right) \tilde{\nabla}_\mu \partial_\nu \Phi = 0 \,,
\end{split}
\end{equation}
which expresses $\tilde{R}_{\mu \nu}$ in terms of functions of $\Phi$ and its derivatives. Having established the Scalar-Tensor equivalence, let us observe that one could also have arrived at the same result by directly studying the equations of motion of the Scalar-Tensor action \eqref{equivfin} with potential $V(\Psi)=-\frac{\Psi^{2}}{{4}}$.\footnote{This follows immediately from the defining relation of $V(\Psi)$ and for the choice $f(R)=R^{2}$. Here there also is another way to arrive at the same result: Integrating $V'=-\chi$ we get $V(\Psi)=-\int \chi{(\Psi)}d\Psi$; then, using the {fact that from \eqref{fsecRchi} we have, for $f''(\chi)\neq 0$, $\chi = R$, here we find (using also \eqref{PsiPhi} and \eqref{goodform}) $\chi=\frac{\Psi}{2}$, which in turn} implies $V(\Psi)=-\frac{\Psi^{2}}{{4}}$. In this way, taking the classical mechanics analogue too far, one might also interpret $\chi(\Psi)$ as the conservative `force' which, when integrated, gives us the potential $V(\Psi)$.} {Indeed, one can prove that the theory is on-shell equivalent to a Scalar-Tensor model with the aforementioned potential.} 

We will now give an application {in the cosmology context, showing, in particular,} how \eqref{dif} may be solved to obtain the scalar curvature $R${.}

\subsubsection{Homogeneous cosmology}

Let us now consider an illuminating application in order to illustrate how \eqref{dif} may be solved to obtain $R$ {in a cosmology context}. In a homogeneous cosmological background we have $\partial_{\mu}\Phi=\delta_{\mu}^{0}\dot{\Phi}$, and the equation we have to solve{, that is \eqref{dif},} becomes (here, $a(t)$ is the scale factor)
\beq
\frac{d}{dt}(a^{3}\dot{\Phi})+a^{3}\dot{\Phi}^{2}\tan{\Phi}=0 \,.
\eeq
If we now express the latter as
\beq
\frac{1}{a^{3}\dot{\Phi}}\frac{d}{dt}(a^{3}\dot{\Phi})=-\left(\tan{\Phi} \right)\dot{\Phi} \,,
\eeq
we can see that {it can be trivially integrated} to get
\beq\label{a3dotPhi}
a^{3}\dot{\Phi}=\cos{\Phi}+c_{1} \,,
\eeq
where $c_{1}$ is an arbitrary integration constant. For the sake of simplicity, let us assume that the initial conditions are such that $c_{1}=0$. Then, integrating \eqref{a3dotPhi}, we arrive at
\beq
\Phi(t)=\arcsin{\left(  \frac{Ce^{\lambda}-1}{C e^{\lambda}+1}    \right)} \,,
\eeq
where $C$ is another integration constant and where we have defined
\beq
\lambda(t):=2 \int \frac{dt}{a^{3}(t)} \,. \label{lambdadef}
\eeq
Thus, using the above results, eq. \eqref{goodform} yields
\beq
R(t)=2\alpha \tan{\left[ \arcsin{\left(  \frac{Ce^{{\lambda}}-1}{C e^{{\lambda}}+1}    \right)} \right]} \,.
\eeq
{Note also that the latter can be also formally rewritten as
\begin{equation}
R(t)= \frac{2\alpha \varphi}{\sqrt{1-\varphi^2}} \,, 
\end{equation}
where we have defined
\begin{equation}
\varphi := \frac{Ce^{\lambda}-1}{C e^{\lambda}+1} \,,
\end{equation}
with $\lambda$ being defined in \eqref{lambdadef}.}
Then, exploiting the decomposition of $R$ in terms of $\tilde{R}$ plus non-Riemannian contributions, one gets an evolution equation for the scale factor, which, in this case, is an integral equation.

\section{Conclusions}\label{conclusions}

Considering a {Metric-Affine} set-up, we have studied a gravitational theory given by the EH term along with a parity violating term defined by the contraction of the Riemann tensor with the Levi-Civita tensor. The latter is oftentimes referred to as the Holst term. Interestingly, in the MAG framework, this is the most general theory one can write down whose gravitational part is linear in the Riemann tensor. Starting with the vacuum case, we have shown that the phenomenology of the model depends crucially on the values of the dimensionless coupling constant of the Holst term. 
In particular, there are two distinct cases to study separately: $\alpha \neq \pm \frac{i}{2}$ and  $\alpha=\pm \frac{i}{2}$. In the former case, {projective invariance of the action allows to appropriately choose the gauge} in such a way to end up with a final connection that is purely Levi-Civita and prove that the theory is equivalent to vacuum GR. On the other hand, in the latter case ($\alpha=\pm \frac{i}{2}$), we have found that, intriguingly, the total action EH + Holst enjoys an enlarged symmetry, namely it is invariant under connection transformations given by \eqref{enlsymm}. Thus, the addition of the Holst term not only does not break the projective invariance of the theory {(since itself respects this symmetry)} but for $\alpha=\pm \frac{i}{2}$ it promotes the projective group to a greater one. {We have then proved that exploiting the invariances of the action, also in the case $\alpha=\pm \frac{i}{2}$ the final connection coincides with the Levi-Civita one and the theory results to be equivalent to GR in vacuum.} Note that neither the EH {nor} the Holst term is invariant under \eqref{enlsymm}, but their linear combination for $\alpha=\pm \frac{i}{2}$ is indeed left invariant. To our knowledge, this invariance is reported for the first time here and it restricts the matter form that can be coupled to the theory. To show this explicitly, we have also added matter to the model and derived the aforementioned restriction imposed by {the enlarged} symmetry on the hypermomentum.

Subsequently, we have extended our discussion and studied the {Metric-Affine} version of $f(R)$ + Holst term gravity. In this case, we have demonstrated how one can always consistently set the nonmetricity to zero by {exploiting} the projective freedom appropriately. The whole analysis then boils down to the two distinctive cases $f'=\text{constant}$ and $f' \neq \text{constant}$. In the former case, we have $f(R)=C_{0}R+C_{1}$, which means that the theory becomes EH + Holst {+} cosmological constant. Then, the results are identical to the EH + Holst theory since the cosmological constant does not modify anything apart from adding an extra term to the metric field equations. Far more interesting is the other possibility, namely $f' \neq \text{constant}$. In this case, as we have explicitly proved, the theory is on-shell equivalent to a metric-compatible torsionless Scalar-Tensor theory, that is we have arrived at the following remarkable result: Vacuum {Metric-Affine} $f(R)$ + Holst term gravity is on-shell equivalent to a metric torsionless Scalar-Tensor theory. In addition, for $\alpha << 1$ the aforementioned Scalar-Tensor theory actually reduces to a Brans-Dicke theory with {Brans-Dicke parameter} $\omega_{0}=-\frac{3}{2}$.

A future development could consist in studying how matter with vanishing hypermomentum affects the results we have obtained in the case of {Metric-Affine} $f(R)$ gravity plus Holst term (and its on-shell equivalence with a metric torsionless Scalar-Tensor theory). 
It would be worth to extend our analysis to the case in which the BI parameter is promoted to a field [work in progress], on the same lines of \cite{TorresGomez:2008fj,Bombacigno:2016siz,Bombacigno:2018tbo,Bombacigno:2018tih}, considering separately the addition of the Holst and of the Nieh-Yan terms in this case. {Indeed, the Nieh-Yan term, which is included into the Holst one, has been proved to provide many remarkable applications, among which the ones discussed in \cite{Langvik:2020nrs,leigh2009torsion,petkou2010torsional} and also \cite{Bombacigno:2018tyw}.} Furthermore, applications to the cosmological perfect hyperfluid of \cite{Iosifidis:2020gth} deserve to be analyzed and discussed.
It would be also interesting to carry on an analysis on the same lines of what we have done in the present paper but considering {a} more general theory developed in \cite{DAuria:1981ddz} in a {Metric-Affine} framework, that is involving both torsion and nonmetricity from the very beginning. In particular, in \cite{DAuria:1981ddz} the authors considered the EH action plus Holst term with torsion in first order formalism and then also generalized their scheme upon the introduction of other geometrical
terms, studying instantonic solutions of the theory. There, some surprising features and analogies with superconductivity arose, obtaining topological singularities which are different
from those of pure gravity. The gravity theory presented in \cite{DAuria:1981ddz} does not imply any asymptotic statement on spacetime and its effect are expected
to be important for small scale phenomena, and it would be therefore intriguing to study possible generalizations of these results in the {Metric-Affine} set-up.

\section*{Acknowledgements}

L.R. would like to thank A. Gamba and F. Dolcini for financial support.


\bibliographystyle{unsrt}
 
\bibliography{references}

\begin{thebibliography}{10}

\bibitem{eisenhart2012non}
Luther~Pfahler Eisenhart.
\newblock {\em Non-riemannian geometry}.
\newblock Courier Corporation, 2012.

\bibitem{Hehl:1994ue}
Friedrich~W. Hehl, J.Dermott McCrea, Eckehard~W. Mielke, and Yuval Ne'eman.
\newblock {Metric affine gauge theory of gravity: Field equations, Noether
  identities, world spinors, and breaking of dilation invariance}.
\newblock {\em Phys. Rept.}, 258:1--171, 1995.

\bibitem{Iosifidis:2019jgi}
Damianos Iosifidis.
\newblock {Metric-Affine Gravity and Cosmology/Aspects of Torsion and
  non-Metricity in Gravity Theories}.
\newblock Other thesis, 2 2019.

\bibitem{Sotiriou:2008rp}
Thomas~P. Sotiriou and Valerio Faraoni.
\newblock {f(R) Theories Of Gravity}.
\newblock {\em Rev. Mod. Phys.}, 82:451--497, 2010.

\bibitem{teyssandier1983cauchy}
Pierre Teyssandier and Ph~Tourrenc.
\newblock The cauchy problem for the r+ r 2 theories of gravity without
  torsion.
\newblock {\em Journal of mathematical physics}, 24(12):2793--2799, 1983.

\bibitem{sotiriou2006f}
Thomas~P Sotiriou.
\newblock f (r) gravity and scalar--tensor theory.
\newblock {\em Classical and Quantum Gravity}, 23(17):5117, 2006.

\bibitem{Klemm:2018bil}
Dietmar~Silke Klemm and Lucrezia Ravera.
\newblock {Einstein manifolds with torsion and nonmetricity}.
\newblock {\em Phys. Rev. D}, 101(4):044011, 2020.

\bibitem{Klemm:2020mfp}
Silke Klemm and Lucrezia Ravera.
\newblock {An action principle for the Einstein-Weyl equations}.
\newblock 6 2020.

\bibitem{Klemm:2019izb}
Dietmar~Silke Klemm and Lucrezia Ravera.
\newblock {Supersymmetric near-horizon geometry and Einstein-Cartan-Weyl
  spaces}.
\newblock {\em Phys. Lett. B}, 793:265--270, 2019.

\bibitem{Klemm:2020gfm}
Silke Klemm and Lucrezia Ravera.
\newblock {Schr\"odinger connection with selfdual nonmetricity vector in 2+1
  dimensions}.
\newblock 8 2020.

\bibitem{Holst:1995pc}
Soren Holst.
\newblock {Barbero's Hamiltonian derived from a generalized Hilbert-Palatini
  action}.
\newblock {\em Phys. Rev. D}, 53:5966--5969, 1996.

\bibitem{Kazmierczak:2008iw}
Marcin Kazmierczak.
\newblock {Einstein-Cartan gravity with Holst term and fermions}.
\newblock {\em Phys. Rev. D}, 79:064029, 2009.

\bibitem{Banerjee:2010yn}
Kinjal Banerjee.
\newblock {Some Aspects of Holst and Nieh-Yan Terms in General Relativity with
  Torsion}.
\newblock {\em Class. Quant. Grav.}, 27:135012, 2010.

\bibitem{Shapiro:2014kma}
Ilya~L. Shapiro and Poliane~M. Teixeira.
\newblock {Quantum Einstein-Cartan theory with the Holst term}.
\newblock {\em Class. Quant. Grav.}, 31:185002, 2014.

\bibitem{Barbero:1994ap}
J.Fernando Barbero~G.
\newblock {Real Ashtekar variables for Lorentzian signature space times}.
\newblock {\em Phys. Rev. D}, 51:5507--5510, 1995.

\bibitem{Immirzi:1996di}
Giorgio Immirzi.
\newblock {Real and complex connections for canonical gravity}.
\newblock {\em Class. Quant. Grav.}, 14:L177--L181, 1997.

\bibitem{Geiller:2012dd}
Marc Geiller and Karim Noui.
\newblock {A note on the Holst action, the time gauge, and the Barbero-Immirzi
  parameter}.
\newblock {\em Gen. Rel. Grav.}, 45:1733--1760, 2013.

\bibitem{Iosifidis:2018zjj}
Damianos Iosifidis, Anastasios~C. Petkou, and Christos~G. Tsagas.
\newblock {Torsion/non-metricity duality in f(R) gravity}.
\newblock {\em Gen. Rel. Grav.}, 51(5):66, 2019.

\bibitem{Iosifidis:2019fsh}
Damianos Iosifidis.
\newblock {Linear Transformations on Affine-Connections}.
\newblock {\em Class. Quant. Grav.}, 37(8):085010, 2020.

\bibitem{Samuel:1987td}
J.~Samuel.
\newblock {A Lagrangian basis for Ashtekar's formulation of canonical gravity}.
\newblock {\em Pramana}, 28:L429--L432, 1987.

\bibitem{Jacobson:1987yw}
T.~Jacobson and L.~Smolin.
\newblock {The Left-Handed Spin Connection as a Variable for Canonical
  Gravity}.
\newblock {\em Phys. Lett. B}, 196:39--42, 1987.

\bibitem{Jacobson:1988yy}
Ted Jacobson and Lee Smolin.
\newblock {Covariant Action for Ashtekar's Form of Canonical Gravity}.
\newblock {\em Class. Quant. Grav.}, 5:583, 1988.

\bibitem{Goldberg:1988ir}
J.N. Goldberg.
\newblock {Triad Approach to the Hamiltonian of General Relativity}.
\newblock {\em Phys. Rev. D}, 37:2116--2120, 1988.

\bibitem{Henneaux:1989hf}
M.~Henneaux, C.~Schomblond, and J.E. Nelson.
\newblock {Derivation of Ashtekar Variables From Tetrad Gravity}.
\newblock {\em Phys. Rev. D}, 39:434--437, 1989.

\bibitem{Hehl:1976kv}
F.W. Hehl, G.D. Kerlick, and P.~Von Der~Heyde.
\newblock {On Hypermomentum in General Relativity. 3. Coupling Hypermomentum to
  Geometry}.
\newblock {\em Z. Naturforsch. A}, 31:823--827, 1976.

\bibitem{obukhov1993hyperfluid}
Yuri~N Obukhov and Romualdo Tresguerres.
\newblock Hyperfluid—a model of classical matter with hypermomentum.
\newblock {\em Physics Letters A}, 184(1):17--22, 1993.

\bibitem{obukhov1996model}
Yuri~N Obukhov.
\newblock On a model of an unconstrained hyperfluid.
\newblock {\em Physics Letters A}, 210(3):163--167, 1996.

\bibitem{Iosifidis:2020gth}
Damianos Iosifidis.
\newblock {Cosmological Hyperfluids, Torsion and Non-metricity}.
\newblock 3 2020.

\bibitem{1981GReGr..13.1037H}
F.~W. {Hehl}, E.~A. {Lord}, and L.~L. {Smalley}.
\newblock {Metric-affine variational principles in general relativity II.
  Relaxation of the Riemannian constraint}.
\newblock {\em General Relativity and Gravitation}, 13:1037--1056, November
  1981.

\bibitem{vitagliano2011dynamics}
Vincenzo Vitagliano, Thomas~P Sotiriou, and Stefano Liberati.
\newblock The dynamics of metric-affine gravity.
\newblock {\em Annals of Physics}, 326(5):1259--1273, 2011.

\bibitem{iosifidis2019exactly}
Damianos Iosifidis.
\newblock Exactly solvable connections in metric-affine gravity.
\newblock {\em Classical and Quantum Gravity}, 36(8):085001, 2019.

\bibitem{jimenez2018teleparallel}
Jose~Beltr{\'a}n Jim{\'e}nez, Lavinia Heisenberg, and Tomi~S Koivisto.
\newblock Teleparallel palatini theories.
\newblock {\em Journal of Cosmology and Astroparticle Physics}, 2018(08):039,
  2018.

\bibitem{jimenez2019ghosts}
Jose~Beltr{\'a}n Jim{\'e}nez and Adria Delhom.
\newblock Ghosts in metric-affine higher order curvature gravity.
\newblock {\em The European Physical Journal C}, 79(8):656, 2019.

\bibitem{aoki2019scalar}
Katsuki Aoki and Keigo Shimada.
\newblock Scalar-metric-affine theories: Can we get ghost-free theories from
  symmetry?
\newblock {\em Physical Review D}, 100(4):044037, 2019.

\bibitem{percacci2020new}
R~Percacci and E~Sezgin.
\newblock New class of ghost-and tachyon-free metric affine gravities.
\newblock {\em Physical Review D}, 101(8):084040, 2020.

\bibitem{Shaposhnikov:2020aen}
Mikhail Shaposhnikov, Andrey Shkerin, Inar Timiryasov, and Sebastian Zell.
\newblock {Einstein-Cartan Portal to Dark Matter}.
\newblock 8 2020.

\bibitem{sotiriou2007metric}
Thomas~P Sotiriou and Stefano Liberati.
\newblock Metric-affine f (r) theories of gravity.
\newblock {\em Annals of Physics}, 322(4):935--966, 2007.

\bibitem{Bombacigno:2018tih}
Flavio Bombacigno and Giovanni Montani.
\newblock {Implications of the Holst term in a $f(R)$ theory with torsion}.
\newblock {\em Phys. Rev. D}, 99(6):064016, 2019.

\bibitem{TorresGomez:2008fj}
Alexander Torres-Gomez and Kirill Krasnov.
\newblock {Remarks on Barbero-Immirzi parameter as a field}.
\newblock {\em Phys. Rev. D}, 79:104014, 2009.

\bibitem{Bombacigno:2016siz}
Flavio Bombacigno, Francesco Cianfrani, and Giovanni Montani.
\newblock {Big-Bounce cosmology in the presence of Immirzi field}.
\newblock {\em Phys. Rev. D}, 94(6):064021, 2016.

\bibitem{Bombacigno:2018tbo}
Flavio Bombacigno and Giovanni Montani.
\newblock {$f(R)$ gravity with Torsion and the Immirzi field: signature for GW
  detection}.
\newblock {\em Phys. Rev. D}, 97:124066, 2019.

\bibitem{Langvik:2020nrs}
Miklos Långvik, Juha-Matti Ojanperä, Sami Raatikainen, and Syksy Rasanen.
\newblock {Higgs inflation with the Holst and the Nieh-Yan term}.
\newblock 7 2020.

\bibitem{leigh2009torsion}
Robert~G Leigh, Nam~Nguyen Hoang, and Anastasios~C Petkou.
\newblock Torsion and the gravity dual of parity breaking in ads4/cft3
  holography.
\newblock {\em Journal of High Energy Physics}, 2009(03):033, 2009.

\bibitem{petkou2010torsional}
Anastasios~C Petkou.
\newblock Torsional degrees of freedom in ads4/cft3.
\newblock {\em arXiv preprint arXiv:1004.1640}, 2010.

\bibitem{Bombacigno:2018tyw}
Flavio Bombacigno and Giovanni Montani.
\newblock {Big bounce cosmology for Palatini $R^2$ gravity with a Nieh--Yan
  term}.
\newblock {\em Eur. Phys. J. C}, 79(5):405, 2019.

\bibitem{DAuria:1981ddz}
R.~D'Auria and T.~Regge.
\newblock {Gravity Theories With Asymptotically Flat Instantons}.
\newblock {\em Nucl. Phys. B}, 195:308--324, 1982.

\end{thebibliography}

\end{document}